\documentclass[fleqn,usenatbib]{mnras}
\usepackage{enumerate}
\usepackage{tikz}
\usepackage{amsmath}
\usepackage{amssymb}
\usepackage{commath}
\usepackage{bbold}

\usepackage{newtxtext,newtxmath}

\usepackage{hhline}
\usepackage{array}
\newcolumntype{P}[1]{>{\centering\arraybackslash}p{#1}}
\usepackage{color}

\definecolor{steelblue}{RGB}{25,25,112}
\definecolor{dullblue}{rgb}{0,0.298,0.49}
\definecolor{darkred}{rgb}{0.545,0,0}
\definecolor{blue2}{cmyk}{1, 0.1, 0.1, 0}

\usepackage{booktabs}

\usepackage{dsfont}
\usepackage{hyperref}
\hypersetup{linktocpage = true}
\usetikzlibrary{arrows,positioning}

\usepackage{graphicx}
\usepackage{dcolumn}
\usepackage{bm}
\def\Msun{{\rm M}_\odot}

\usepackage{fontawesome}
\usepackage{chemformula}
\usepackage[utf8]{inputenc}

\definecolor{steelblue}{RGB}{25,25,112}

\hypersetup{colorlinks,linkcolor={darkred},citecolor={dullblue},urlcolor={dullblue}}  

\usepackage{colortbl}
\definecolor{blue2}{cmyk}{1, 0.1, 0.1, 0}
\definecolor{Green}{RGB}{0, 128, 0}

\usepackage[T1]{fontenc}
\usepackage{ae,aecompl}

\title[Non-Universal Stellar Initial Mass Functions]{Non-Universal Stellar Initial Mass Functions: Large Uncertainties in Star Formation Rates at $z \approx 2-4$ and Other Astrophysical Probes.}

\author[J. J. Ziegler et al.]{
Joshua J. Ziegler,$^{1}$
\thanks{jjziegler@utexas.edu}
Thomas D. P. Edwards,$^{2,3,4}$
\thanks{thomas.edwards@fysik.su.se}
Anna M. Suliga,$^{5,6,7}$
\thanks{asuliga@berkeley.edu}
Irene Tamborra,$^{7}$
\thanks{tamborra@nbi.ku.dk}
\newauthor
Shunsaku Horiuchi,$^{8,9}$
\thanks{horiuchi@vt.edu}
Shin'ichiro Ando,$^{2,9}$
\thanks{s.ando@uva.nl}
and Katherine Freese$^{1,3,4}$
\thanks{ktfreese@utexas.edu}
\\
$^{1}$Department of Physics, Department of Physics, Austin, Texas 78751, USA\\
$^{2}$Gravitation Astroparticle Physics Amsterdam (GRAPPA), Institute for Theoretical Physics Amsterdam and Delta Institute for Theoretical Physics,\\ University of Amsterdam, Science Park 904, 1090 GL Amsterdam, The Netherlands\\
$^{3}$The Oskar Klein Centre, Department of Physics, Stockholm University, AlbaNova, SE-10691 Stockholm, Sweden\\
$^{4}$Nordic Institute for Theoretical Physics (NORDITA), 106 91 Stockholm, Sweden\\
$^{5}$Department of Physics, University of California Berkeley, Berkeley, California 94720, USA\\
$^{6}$Department of Physics, University of Wisconsin-Madison,
Madison, Wisconsin 53706, USA\\
$^{7}$Niels Bohr International Academy and DARK, Niels Bohr Institute, University of Copenhagen, Blegdamsvej 17, 2100, Copenhagen, Denmark\\
$^{8}$Center for Neutrino Physics, Department of Physics,
Virginia Tech, Blacksburg, Virginia 24061, USA\\
$^{9}$Kavli IPMU (WPI), UTIAS, The University of Tokyo, Kashiwa, Chiba 277-8583, Japan
}

\date{Accepted XXX. Received YYY; in original form ZZZ}

\pubyear{2022}

\begin{document}

\label{firstpage}
\pagerange{\pageref{firstpage}--\pageref{lastpage}}

\maketitle

\begin{abstract}
We explore the assumption, widely used in many astrophysical calculations, that the stellar initial mass function (IMF) is universal across all galaxies. By considering both a canonical Salpeter-like IMF and a non-universal IMF, we are able to compare the effect of different IMFs on multiple observables and derived quantities in astrophysics. Specifically, we consider a non-universal IMF which varies as a function of the local star formation rate, and explore the effects on the star formation rate density (SFRD), the extragalactic background light, the supernova (both core-collapse and thermonuclear) rates, and the diffuse supernova neutrino background.
Our most interesting result is that our adopted varying IMF leads to much greater uncertainty on the SFRD at $z \approx 2-4$ than is usually assumed.  Indeed, we find a SFRD (inferred using observed galaxy luminosity distributions) that is a factor of $\gtrsim 3$ lower than canonical results obtained using a universal Salpeter-like IMF. 
Secondly, the non-universal IMF we explore implies a reduction in the supernova core-collapse rate of a factor of $\sim2$, compared against a universal IMF.
The other potential tracers are only slightly affected by changes to the properties of the IMF. We find that currently available data do not provide a clear preference for universal or non-universal IMF.
However, improvements to measurements of the star formation rate and core-collapse supernova rate at redshifts $z \gtrsim 2$ may offer the best prospects for discernment.

\end{abstract}



\maketitle

\section{Introduction}

In order to understand the formation and evolution of stars, an important quantity is the stellar initial mass function (IMF), the relative numbers of stars as a function of their mass at the time of their formation.  As yet, the IMF remains only loosely constrained observationally. A common assumption is that the IMF is universal -- the same in all environments and throughout cosmic time.  In this paper, we examine five observables that vary over cosmological distances and which strongly depend on the high-mass region of the IMF. One of our goals is to identify the extent to which these observables can be used to test the assumption of a universal IMF at the high-mass end. In particular, we study the consequences of non-universal IMFs for various astrophysical quantities, finding larger uncertainties in the star formation rate and the core-collapse supernova rate.

The concept of an IMF was introduced by \citet{1955ApJ...121..161S}, who proposed 
a single power law 
$\frac{dN}{dM} \propto M^{\alpha}\,$ where $N$ is the number of stars formed with mass $M$; in what is now known as the Salpeter IMF,
he took $\alpha = - 2.35$.  With the assumption of a single power law, the exponent $\alpha$ can be measured to within approximately 10\%~\citep{2003ApJ...593..258B}. Unfortunately, there are fundamental questions about the parametrization that should be used in describing the IMF. Perhaps most notably, it was recognized at the end of the 20th century that low mass stars did not tend to fall on the power-law distribution predicted by Salpeter. This gave rise to IMF models with low-mass suppressions, such as the broken power law of \citet{2001MNRAS.322..231K} and the log-normal distribution of \citet{2003PASP..115..763C}.
Recent evidence suggests that the IMF may even have an intrinsic dependence on the local environment~\citep{2010Natur.468..940V, FM2013, 2012Natur.484..485C, Ferreras2012, LaBarbera2019, Harayama2008, 2011MNRAS.415.1647G}. Variations to the low-mass end of the IMF have been studied extensively in the literature~\citep{2010Natur.468..940V, 2003PASP..115..763C, 2013hst..prop13449G}. Instead, following recent evidence~\citep{2011MNRAS.415.1647G}, we focus on observables sensitive to the high-mass end of the IMF which may also be non-universal.

Star-forming regions can be distinguished by a variety of properties of the collapsing gas and dust, including angular momentum, metallicity, density, temperature, and dust content. 
The universality of the IMF therefore boils down to an assumption that all of these properties play little to no role in the masses of the formed stars. 
Whether this is theoretically justified remains unclear.
As described in \citet{Offner2014} (and references therein), perturbations in the density of a star-forming gas cloud can, under reasonable assumptions, generate a power-law spectrum of core and clump masses, where cores and clumps refer to gravitationally collapsing gas clouds that are likely to form at least one star. In contrast to this power-law distribution, at low masses, turbulence in the star-forming cloud can naturally produce a spectrum of masses which disfavors lower mass stars relative to the power-law predictions. In particular, \citet{1997MNRAS.288..145P} showed that turbulence could give rise to a log-normal mass distribution among low mass cores/clumps, similar to the IMF described by \citet{2003PASP..115..763C}. While this theoretical explanation would seem to leave very little room for non-universality in the IMF, the mass function described here is for cores and clumps, not stars. In relating this mass function to the stellar IMF, numerous assumptions must be made about the formation of protostars out of collapsing gas \citep{Offner2014}. The validity of many of these assumptions, especially in extreme environments, is largely an open question, suggesting that even within this theoretical framework, there may be room to consider non-universality without requiring a new paradigm.

The question of whether the IMF is indeed universal has been investigated many times. For example, despite most observations being consistent with a universal IMF, authors have regularly suggested a non-universal IMF as a way to explain other astrophysical tensions~\citep{1998MNRAS.301..569L}. Further, over the last two decades, hints of a tension between universal IMFs and observations have developed, particularly in early-type elliptical galaxies~\citep{2010Natur.468..940V, FM2013, 2012Natur.484..485C, Ferreras2012, LaBarbera2019} and in environments which experience extreme properties~\citep{Harayama2008, 2011MNRAS.415.1647G}. Theoretical models, such as the Integrated Galaxy-wide IMF~\citep{2017MNRAS.464.3812F} and the cosmic-ray-regulated star formation discussed in \citet{Fontanot_2018}, can offer justifications for some of these observations and pose additional predictions. On the other hand, due to the inherent difficulty in measuring the IMF, many authors reject these observational claims, leaving the question of whether the IMF is indeed universal largely unanswered (see e.g., \citet{Hopkins2018}, and references within for a recent review of the range of perspectives).

A large part of the uncertainty in whether the IMF is universal can be traced to the difficulty in unambiguously measuring it. Locally, where it is possible to resolve individual stars, one can estimate the IMF by comparing observed stellar populations to the populations that are predicted to form if different IMF models are assumed~\citep{2013seg..book..419C, KE2012}. While accurate, this approach can only be used in star-forming regions nearby enough to resolve individual stars, and requires assumptions about the history of star formation in that region. On the other hand, for more distant galaxies, where it is impossible to see inside star-forming regions or where it is difficult to resolve individual stars, some proxy for the star formation rate (SFR) must be used. The most common approach is to use the luminosity as a measure of the rate of star formation (for example \citet{1998ARA&A..36..189K}), but this causes observations of the IMF to depend heavily on the calibration factor between luminosity and SFR. While it is possible to calculate this calibration factor numerically, it requires an assumption about the IMF. Unfortunately, this circular dependence encountered when calculating the IMF of distant galaxies is rather ubiquitous, making an independent measurement of the IMF challenging.

While it is difficult to directly measure the IMF, it may be possible to find indirect ways to probe the effects of a non-universal IMF. Previous works, such as \citet{2017MNRAS.464.3812F,Fontanot_2018} approach this problem as well, but use different models of varying IMF and probe different astrophysical observables than we do here. 

In this paper, we examine five observables that vary over cosmological distances and which depend on the IMF:  the star formation rate density (SFRD), the extragalactic background light (EBL), the core-collapse supernova (CCSN) rate density, the type Ia supernova (SNIa) rate, and the diffuse supernova neutrino background (DSNB). For each, we explore how they change when using a non-universal IMF compared to a universal one, and discuss whether they are discriminable with current or future data. For simplicity, we focus on the change induced by a varying IMF and ignore many other uncertainties directly related to each observable. These additional uncertainties will, in practice, make it more difficult to observe the IMF induced changes. Our goal is simply to learn whether astrophysical observations of distant objects could, in principle, provide indirect evidence for a non-universal IMF.

The rest of this paper is structured as follows. In Sec.~\ref{sec:varIMF}, we describe the two IMF models we consider throughout the paper. In Sec.~\ref{implications}, we look at how these IMF models affect the five quantities described in the previous paragraph: the star formation rate density (SFRD), the extragalactic background light (EBL), the core-collapse supernova (CCSN) rate density, the type Ia supernova (SNIa) rate, and the diffuse supernova neutrino background (DSNB). Finally, we conclude in Sec.~\ref{sec:conclusion}.

\section{Initial Mass Function Models}
\label{sec:varIMF}

In a star-forming region, the stellar IMF describes the distribution of masses with which stars form. A common approach to describing this IMF is through a probability distribution $\xi(M)$. That is
\begin{equation}
    \frac{dN}{dM} = \xi(M)N_{\rm tot}\,, 
\end{equation}
where $N$ is the number of stars formed with mass $M$, typically measured in units of $\Msun$, and $N_{\rm tot}$ is the total number of stars formed. Under this convention, $\xi(M)$ is normalized such that $\int{\xi(M)dM} = 1$, when integrating over all possible stellar masses.

\begin{figure}
    \centering
    \includegraphics[width=\columnwidth]{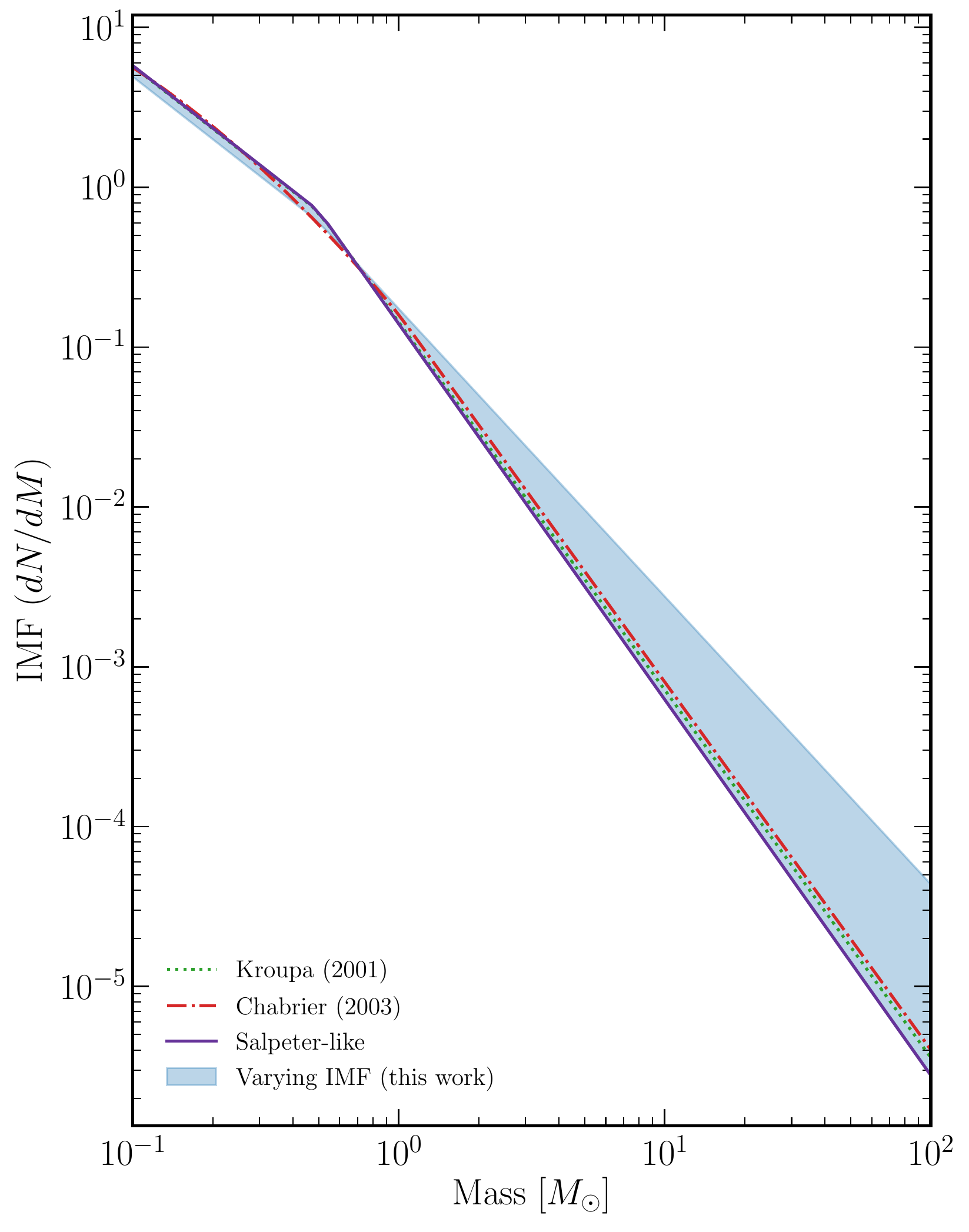}
    \caption{\textbf{Initial mass functions:} This figure compares the non-universal IMFs we use in this paper (blue shaded region) to the canonical IMFs used in the literature:
    \citet{2001MNRAS.322..231K} (green dotted) and \citet{2003PASP..115..763C} (red dash-dotted).  We note the similarity of the two canonical cases relative to the wide range of non-universal IMFs we consider; we will use a proxy we name the \textit{Salpeter-like} IMF (purple solid line) as our benchmark for the canonical cases. Each IMF is normalized so that the integral over mass equals 1. The Kroupa IMF follows a broken power law, with slope $\alpha = -1.3$ for $M<0.5\,\Msun$ and $\alpha = -2.3$ for $M>0.5\,\Msun$. The Chabrier IMF also behaves differently at low mass versus high mass stars, with the low mass stars ($M< 1\,\Msun$) following a log-normal behavior while high mass stars follow a power law with slope $\alpha = -2.35$. In all cases we use broken power-law IMFs, defined piecewise with a break at 0.5 $\Msun$. \textit{Non-universal IMF} (blue-shaded region): For masses $M<0.5\,\Msun, \alpha=-1.3$ and for $M>0.5\,\Msun, \alpha$ can take values in the range -1.8 to -2.35, shown here in the blue shaded region. \textit{Benchmark universal Salpeter-like IMF} (purple solid curve):  consists of a shallow power law like that in the Kroupa IMF at low masses and a Salpeter $\alpha = -2.35$ IMF at high masses. Throughout this paper, we will use this case as a benchmark against which to test the effects of allowing the IMF to vary.
    }
    \label{fig:IMF}
\end{figure}

In Salpeter's seminal work~\citep{1955ApJ...121..161S}, the IMF was described as a power law of the form 
\begin{equation}
    \xi(M) \propto M^{\alpha}\,,
\end{equation} 
where $\alpha=-2.35$ was observed for stars in a mass range $0.4$ to $1.0 \,\Msun$. Since then, the range of masses over which the IMF could be determined has vastly increased, but the practice of describing the IMF through a power-law slope has remained. However, as IMFs have been studied, the single straight power law of Salpeter has given way to IMFs with more features. For example, commonly used IMFs include the piecewise power law established by \citet{2001MNRAS.322..231K} (a variant of which was used in for example \citet{2003ApJ...593..258B}) and the  \citet{2003PASP..115..763C} IMF which has a log-normal distribution for stars below approximately $1\,\Msun$ and a power law for stars greater than $1\,\Msun$. These two IMFs are plotted in Fig.~\ref{fig:IMF}. As long as a universal IMF is assumed, the high-mass behavior of the IMF is approximately a power law with $\alpha \approx -2.35$, consistent with a Salpeter IMF~\citep{2003ApJ...593..258B}.\footnote{Note that the typical break points for Kroupa and Chabrier are $0.5\,\Msun$ and $1\,\Msun$ respectively. Because the log-normal mass function smoothly turns over, they end up giving a similar distribution of stellar masses.}

However, while it is broadly accepted that a power-law Salpeter IMF does not hold true at low masses, questions regarding the range of environments over which the Salpeter power law is valid for stars with mass greater than $0.5\,\Msun$ remain significantly disputed. One of the most well-motivated regimes in which deviations from a Salpeter IMF at higher stellar masses could occur is the set of conditions in which Population III stars grow. For example, a top-heavy IMF at early times, which favors a higher average mass for Pop III stars, seems to be preferred by observations~\citep{Sharda2021}. One possible mechanism that could justify this behavior is described in \citet{Sharda2021}, where a change in metallicity can shift the peak mass from approximately $0.5\,\Msun$ for solar metallicities to around $50\,\Msun$ at metallicities of Pop III stars. Furthermore, Pop III stars may exhibit energy production mechanisms inaccessible in Pop I and Pop II stars, as would be the case for Dark Stars (stars made of hydrogen and helium but powered by dark matter)~\citep{freese2016}. In that case, it would be quite surprising if the IMF were to be consistent across all three populations.

In addition, there has been a growing body of evidence that seems to suggest that while the IMF behaves as a power law at high masses, the slope may depend substantially on environmental factors.  Various authors identify multiple factors as possible sources of these deviations away from Salpeter. In addition to metallicity, these include: velocity dispersion~\citep{Ferreras2012, FM2013}; radius and surface mass density~\citep{LaBarbera2019}; and high turbulence~\citep{chabrier2014}. 

In this paper, we focus on a non-universal IMF that varies with the SFR of a star-forming region. Specifically, we explore a relationship that was identified in data from the spectroscopic GAMA survey~\citep{2011MNRAS.413..971D}, as analyzed in \citet{2011MNRAS.415.1647G}. The GAMA survey was undertaken by the Anglo-Australian telescope, which had measured the spectra of 120,000 galaxies at the time \citet{2011MNRAS.415.1647G} did their analysis. It has now taken the spectra of approximately 300,000 galaxies. The analysis in \citet{2011MNRAS.415.1647G} used the emission strength of the H$\alpha$ line as a proxy to calculate the SFR of a galaxy, and then binned galaxies based on that SFR. Using a set of simulated galaxies, a power-law IMF was fit to the observed galaxies in each SFR bin, with the exponent $\alpha$ free to vary. Using these binned galaxies, they found a clear preference for a non-universal IMF, and that the variation could be described by the function, 
$\alpha_G \approx 0.36 \log\langle \mathrm{SFR}\rangle - 2.6.$
Here, the average SFR, $\langle \mathrm{SFR} \rangle$, is measured in units of $\Msun \, \mathrm{yr}^{-1}$.

While this expression is the basis of the varying IMF we consider throughout the rest of this work, it is not in the most convenient form for our purposes. In particular, the independent quantity is the SFR, which is inferred from the luminosity of the H$\alpha$ emission line. To calculate the astrophysical observables discussed below, we will need galaxy luminosity functions up to high redshifts. Unfortunately, the H$\alpha$ emission line is not the ideal tracer of these luminosity functions as dust reprocesses most light emitted by galaxies into the infrared.  On the other hand, galaxy surveys (and therefore galaxy luminosity functions) are more complete and readily available in the far-infrared (FIR) band (i.e., in the wavelength range 8-1000 $\mu m$) up to high redshifts. As a result, although using infrared luminosities can introduce significant uncertainty into the calculated SFRs \citep{Madau:2014bja, Wilkins2019}, it is essential for our calculations below.\footnote{
    Note that \citet{Wilkins2019} found that the precise stellar mass range considered can alter the FIR and H$\alpha$ calibration factors, although the alteration is not necessarily the same between the two frequency bands. We therefore point out that by shifting from H$\alpha$ to FIR, we are introducing additional error on the overall magnitude of each of the observables discussed below.
    }
With this in mind, we convert $\langle \mathrm{SFR} \rangle$ to the FIR luminosity $L_{{\rm FIR}}$, using fixed conversion factors from \citet{1998ARA&A..36..189K}.\footnote{Specifically, we used the relation $\mathrm{SFR}\, (\Msun \, \mathrm{yr}^{-1}) = 4.5 \times 10^{-44} L_{FIR}\, (\mathrm{erg} \, \mathrm{s}^{-1})$\citep{1998ARA&A..36..189K}. We discuss how this value depends on the IMF in Section \ref{sec:calib}.} This mimics the process used by \citet{1998ARA&A..36..189K} in reverse, but implicitly assumes that the SFRs predicted by both tracers (H$\alpha$ luminosity and FIR luminosity) are consistent. Using this procedure, we can rewrite the varying IMF expression from \citet{1998ARA&A..36..189K} as
\begin{equation}
    \alpha_{\mathrm{var}, >0.5} \approx 0.36 \log\langle L_{{\rm FIR}} \rangle - 6.1\,.
    \label{IMF LIR}
\end{equation}

Ultimately, the IMF we consider here is empirically based, so we choose to confine ourselves to the range of IMFs that were observed in the corresponding data. In particular, the analysis in \citet{2011MNRAS.415.1647G} calculated IMFs ranging from $\alpha \approx -2.35$ to $-1.8$, with some populations of galaxies having IMFs as steep as $\alpha \approx -2.5$. We limit ourselves to consider only the range of $\alpha \in [-2.35, -1.8],$ which ensures that low luminosity galaxies have an IMF with slope $\alpha=-2.35$.\footnote{
    We note however, that increasing the range of possible $\alpha$'s does not significantly affect our results. In particular, we verified that extending the range to $\alpha \in [-2.5, -1.8]$ has no noticeable effect on all results shown below.
} 
This $\alpha$ range corresponds to enforcing galaxies with a luminosity $\log\langle L/L_\odot \rangle \gtrsim 12$ to have $\alpha = -1.8$, and for galaxies with luminosity $\log\langle L/L_\odot \rangle \lesssim 10.4$ to have $\alpha = -2.35$. Furthermore, we explore only the effect of varying the IMF above a mass cutoff of $0.5\, \Msun$, which gives comparable low-mass behavior to the Chabrier and Kroupa IMFs. Below this mass cutoff, we use a fixed power law 
\begin{equation}
    \alpha_{\mathrm{var}, <0.5} = -1.3\,,
\end{equation}
which matches the low-mass power law of the Kroupa IMF from 0.1 to 0.5 $\Msun$. 

We plot the range of IMFs that may appear in this luminosity-dependent varying IMF in Fig.~\ref{fig:IMF} (blue band). Special attention is given to the IMF which consists of a shallow power law like that in the Kroupa IMF at low masses and a Salpeter IMF at high masses. Throughout this paper, we will use this \textit{Salpeter-like} IMF (blue line) as a benchmark against which to test the effects of allowing the IMF to vary. One important fact that is readily seen from Fig.~\ref{fig:IMF}, and which has been noted by, for example \citet{Hopkins2018}, is that all of the IMF models we consider are quite similar, with only slight variation between them. However, despite the smallness of these variations, when the different IMF models are used to predict the values of observables, especially those that depend on integration of IMF-dependent quantities, we can see substantial differences appearing between the predictions made under those IMF models.

\subsection{Luminosity to SFR Calibration Factor}
\label{sec:calib}

We are interested in using observables which vary on cosmological scales to probe the IMF, and on those scales directly measuring the IMF is unrealistic. Instead, we will be using luminosity as a proxy for star formation, and by extension the IMF, as described in Eq.~\eqref{IMF LIR}. Because all of the observables we consider are related to the rates at which stars form or die, a necessary factor in their calculation is the calibration factor, which we denote $\chi$, that relates luminosity to SFR. In general, the calibration factor depends on the IMF, and because we are considering a non-universal IMF, we must consider how the $\chi$ will depend on our assumed IMF.

For a Salpeter IMF, multiple calculations of the calibration factor have been performed. In particular, \citet{1998ARA&A..36..189K} found the calibration factor using three different wavelength ranges. While modeling of factors such as dust has improved~\citep{KE2012}, the values in \citet{1998ARA&A..36..189K} are often useful as a benchmark for illustrative purposes. Throughout this paper, we focus on the far infrared (FIR) wavelength range, 8-1000 $\mu$m, for which the value of the calibration factor from \citet{1998ARA&A..36..189K}, assuming a Salpeter IMF, is 
\begin{equation}
    \chi_{\mathrm{K98, FIR}} = 4.5 \times 10^{-44} \, \mathrm{\Msun \, yr^{-1} \, s \, erg^{-1}}\,.
\end{equation}

While the values obtained in \citet{1998ARA&A..36..189K} are derived assuming a Salpeter IMF, and are reasonably consistent with a Salpeter-like IMF (as shown in Fig.~\ref{fig:IMF}), we are interested in IMFs with a range of high-mass behaviors. In order to calculate the impact that changing the IMF has on the calibration factor, we use the code \texttt{P\'egase.3} \citep{2019A&A...623A.143F}, which simulates the radiation spectrum of a galaxy with a set of user-defined inputs (our particular scenarios are provided here: \href{https://github.com/joshziegler/DSNB/tree/main/Pegase}{\texttt{P\'egase} inputs}). 
In each \texttt{P\'egase} simulation, a cloud of gas is converted to stars at a prescribed SFR, and with a prescribed IMF. The stellar spectrum of each star that is formed is computed and allowed to evolve according to stellar evolution processes. These stellar spectra are then summed for the ensemble of stars in the galaxy at each timestep in the simulation, resulting in a galactic spectrum. That galactic spectrum is then adjusted to account for the reprocessing of stellar light by dust. Through this process, we can get a spectrum which may be integrated over various frequency ranges and used to calculate luminosities in different frequency bands.

In our simulations, we consider galaxies in the local universe that have a constant rate of star formation, and with other properties that we allow to vary. Besides allowing for different IMFs, we look at galaxies with two different geometries: disky spiral galaxies and spheroidal starburst galaxies, and with different metallicities and different dust models. In particular, for metallicity we consider a high-metallicity case with $Z = 0.026$ and a low-metallicity case with $Z = 0.013$.\footnote{Note that some work has been done to self-consistently model metallicity evolution together with a varying IMF~\citep{2010AIPC.1240..123K}.}
For the dust models, we consider the dust models of \citet{2004ApJS..152..211Z} (ZWD), and \citet{2001ApJ...554..778L,2001ApJ...548..296W} (LWD). We summarize the combinations of the models we used in Table~\ref{tab:calibdustZ}. 

\begin{figure*}
     \centering
     \includegraphics[width=0.49\linewidth]{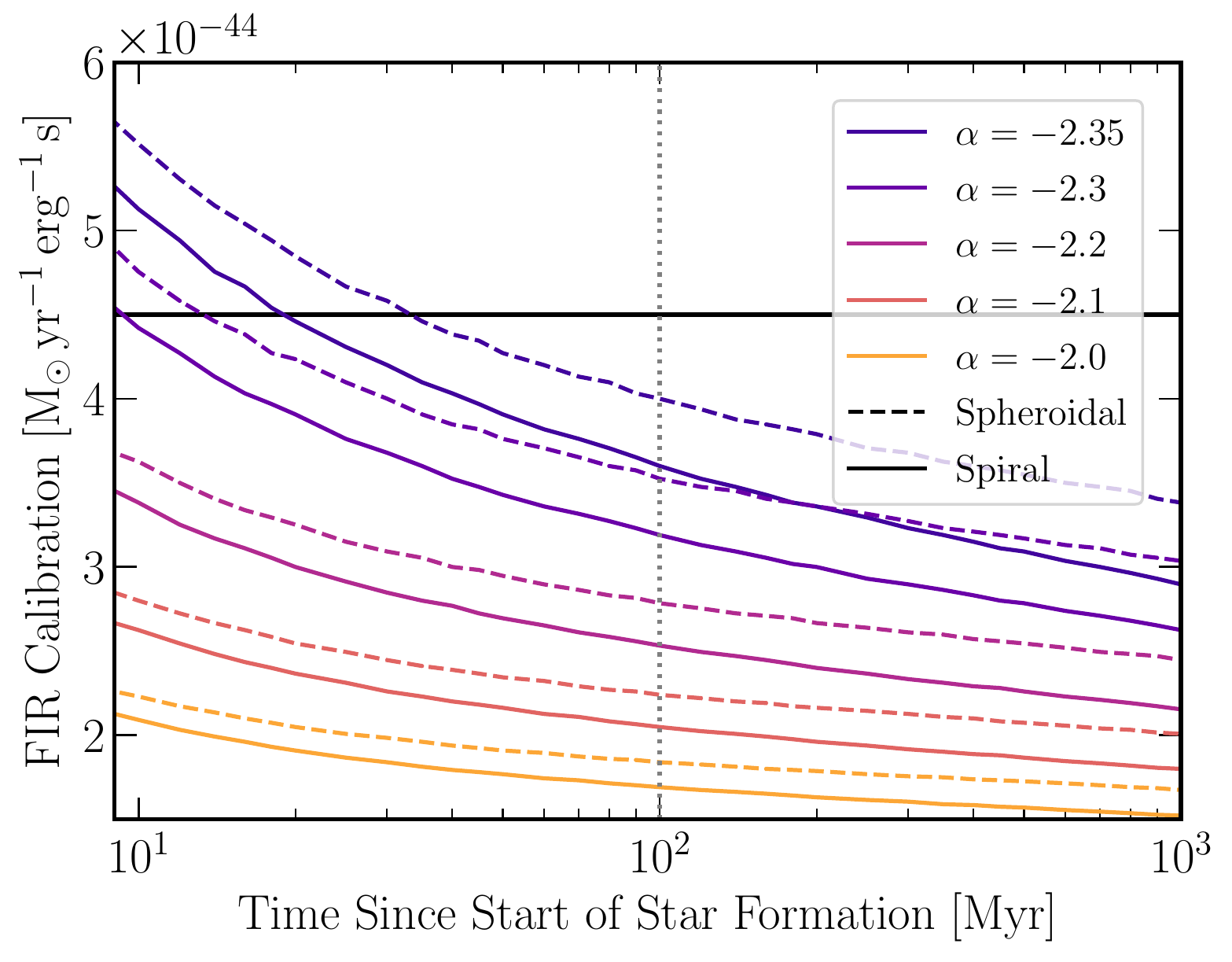}
     \includegraphics[width=0.49\linewidth]{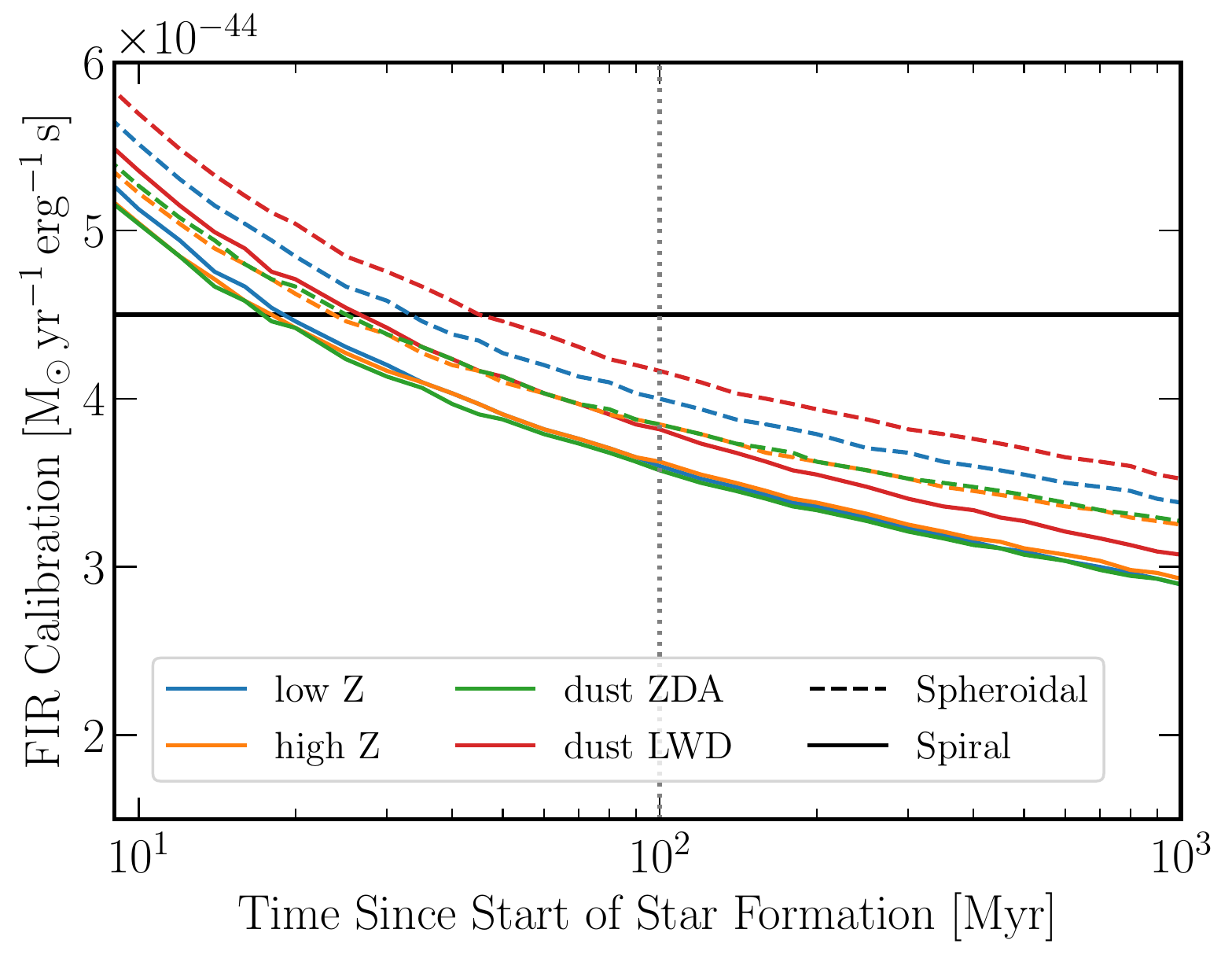}
     \caption{\textbf{Calibration factor vs IMF slope:} \textit{(Left Panel)} Comparison of the calibration factor $\chi = R_{SF}/L$ for a range of IMF slopes ($\alpha = -2.35$ to $-1.8$) used in our \texttt{P\'egase.3} simulations. 
     We also show the impact of galaxy morphology by comparing calibration factors assuming spiral galaxies (solid lines) and spheroidal galaxies (dashed lines). The horizontal black line is the calibration factor identified in Kennicutt 1998~\citep{1998ARA&A..36..189K} for an IMF slope $\alpha = -2.35$ and assuming a spheroidal galaxy. Conventionally, the calibration factor is reported at 100 Myr following an onset of star formation, indicated here by a grey dotted vertical line. \textit{(Right Panel)} The effect of different dust models and metallicities on the calibration factor, for both spiral and spheroidal morphologies with an IMF slope $\alpha = -2.35$. For comparison, all curves in the left panel use the ``low Z'' scenario.}
     \label{fig:calib}
 \end{figure*}

For each of the simulated galaxies, we calculate the luminosity in an $8$--$1000~\mu$m wavelength band at different times from $10$~Myr to $1$~Gyr after the star formation begins. Figure~\ref{fig:calib} shows how the calibration factor is affected considering each of these changes in input parameters. Here it can be clearly seen that while changing the metallicity and dust model does lead to a distinguishable difference in the calibration factor, those differences are small compared to the changes induced by changing the IMF and geometry.

\begin{table}
 \label{tab:calibdustZ}
 \renewcommand{\arraystretch}{1.3}
   \begin{center}
     \begin{tabular}{c|c|c} 
     Scenario & $Z$ & Grains File \\
       \hline
       \hline
       low Z & 0.013 & ZDA\\
       high Z & 0.026  & ZDA\\
       dust ZDA & 0.0195  & ZDA\\
       dust LWD & 0.0195  & LWD\\
     \end{tabular}
     \caption{Choices of input parameters in \texttt{P\'egase.3}. We use the various scenarios to estimate the impact of these parameters on the luminosity to SFR calibration factor. For all other results in the paper, we use the calibration factors derived with the ``low Z'' conditions.}
   \end{center}
 \end{table}

The convention for reporting a single calibration factor is to take the value at 100 Myr after the start of star formation. For other tracers of star formation, particularly ultraviolet luminosity, the calibration factor is effectively a constant after 100 Myr~\citep{1998ARA&A..36..189K}. While that is not the case for the FIR calibration factor, we adopt the same convention. Under this definition, we observe that the calibration factor we calculate for a Salpeter-like IMF and a spheroidal starburst galaxy is within 10\% of the calibration factor determined by Kennicutt~\citep{1998ARA&A..36..189K} for a Salpeter IMF in a spheroidal galaxy. We do not account for other effects, e.g., stellar rotation on the calibrations \citep{Horiuchi:2013bc}. Note that for $\alpha=-2.35$ we compared \texttt{P\'egase} with \texttt{Starburst~99}~\citep{Leitherer:1999rq} and found similar calibration factors.

\section{Probes of a Non-universal IMF}
\label{implications}

Now that we have established our IMF models and the associated calibration factors, we will explore five astrophysical observables that intrinsically depend on the IMF. For each, we present the theoretical prediction for both IMF models and discuss whether current or future data are able to distinguish between the two.

\subsection{Star Formation Rate Density}
\label{sec:SFRD}

We first explore how the SFR of galaxies could provide constraints on the nature of the IMF. The SFR is the rate at which gas in a star-forming region turns into stars, typically measured in $\mathrm{\Msun \, yr^{-1}}$. While an interesting quantity in its own right, we will focus on the related star formation rate density (SFRD), which measures the star formation rate per unit volume and typically has units $\mathrm{\Msun \, yr^{-1} \, Mpc^{-3}}$. By looking at the SFRD rather than individual galaxies' SFR, we can average over the variance introduced because of different galactic properties and specifically explore how star formation depends on redshift. As a result, while both quantities give insight into the star formation process, the SFRD is more directly tied to the cosmic star formation history and less dependent on conditions within individual star-forming regions~\citep{Madau:2014bja}.  

While an understanding of the SFR is critical to theories of galactic evolution, it is challenging to measure directly. In fact, only in local systems, where stars can be resolved, can the SFR be directly measured~\citep{2013seg..book..419C, KE2012}. Where young stars can be resolved, namely within the Milky Way and the nearest galaxies, it is possible to count those young stars and therefore directly estimate the SFR~\citep{2011AJ....142..197C}. In systems slightly more distant, where it is possible to resolve stars but impossible to see young stars shrouded in dust, fitting the galactic color-magnitude diagram to simulations can provide an accurate measure of the SFR, among other properties. However, for more distant systems, in which stars cannot individually be resolved, an indirect measure of the SFR must be used, typically treating luminosity as a tracer of the SFR.\footnote{While the methods described here are among the most direct ways to estimate the SFR, work has been done to improve these estimates by combining these methods with observations that depend indirectly on the SFR. For example, see \citet{2008MNRAS.385..687W,2008MNRAS.391..363W}}
To convert from luminosity to SFR, one uses the calibration factor introduced in the previous section. As mentioned before, unfortunately these calibration factors are determined using simulations which require assumptions to be made about the IMF. In section \ref{sec:calib}, we describe the impact that allowing the IMF to vary can have on the calibration factor.

To calculate the SFRD, we need the calibration factor as well as the luminosity distribution of galaxies. This distribution can be described through the luminosity function ${dN}/{d\log L}$,\footnote{
    Note that all equations below are written for a generic luminosity $L$. For notational simplicity, we therefore drop the subscript FIR on all luminosities. However, all calculations are performed within the FIR luminosity range (8-1000$\,\mu$m).
}
which will generically be a function of redshift $z$. In all of our calculations, we use the set of luminosity functions calculated from data collected by the Herschel observatory in multiple complementary surveys including the PACS Evolutionary Probe (PEP), Herschel Multi-Tiered Extragalactic Survey (HerMES), and Herschel Great Observatories Origin Deep Survey (GOODS) \citep{Gruppioni:2013jna}. By including deep, pencil beam surveys like GOODS, these data include galaxies out to a redshift of $z\sim 4$. Meanwhile, broad, shallow surveys, like those in PEP and HerMES, can help provide more accurate identification of galaxies' morphologies, 
Therefore, from this combined dataset, \citet{Gruppioni:2013jna} were able to develop accurate, galaxy-morphology specific luminosity functions, labeled as ``spiral'', ``starburst'', and ``AGN-SF'' for redshifts $z\approx0-4$.\footnote{
    Note that care must be taken in how AGN are considered. The luminosity functions we consider are derived from the total luminosity of the galaxies, which includes both AGN and stellar sources. However, when calculating calibration factors, it is conventional to include only luminosity from stellar sources, not including AGN. As a result, introducing AGN may decrease the calibration factors from what we derive in the previous section, but we do not consider the effects of such a decrease in this work.
} 
Respectively, these describe: spiral, disky galaxies without extreme star formation; spheroidal galaxies with intense star formation; and galaxies with a bright active galactic nucleus (AGN). We further distinguish the AGN category into spiral galaxies with an AGN and starburst galaxies with anAGN, based on the fraction of each type presented in \citet{Gruppioni:2013jna}. For both spiral galaxies with or without an AGN, we use the spiral calibration factors from the previous section. Similarly, for starburst galaxies with or without an AGN, we use the spheroidal calbiration factors.

\begin{figure}
    \centering
    \includegraphics[width=\columnwidth]{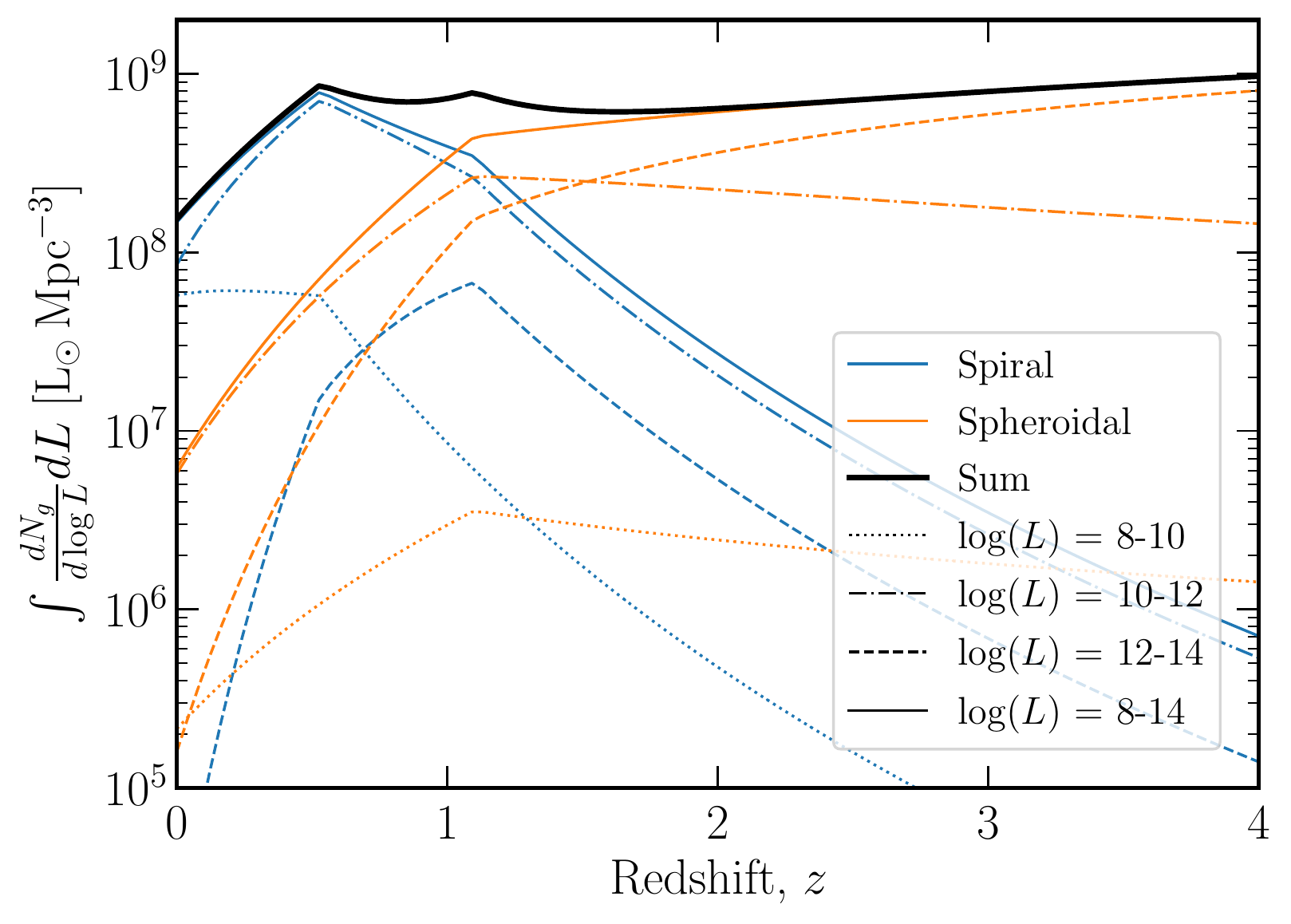}
    \caption{\textbf{Luminosity density:} We show how the luminosity functions ${dN_g}/{d\log L}$ from \citet{Gruppioni:2013jna} depend on redshift. These luminosity functions give the density of galaxies as a function of luminosity and redshift. Here we present the integral $\int dL \: {dN_g}/{d\log L}$ for spiral and starburst galaxies. We also show the contribution to these integrals for the luminosity ranges: $10^8$--$10^{10} L_\odot$, $10^{10}$--$10^{12} L_\odot$, and $10^{12}$--$10^{14} L_\odot$. The luminosity functions in \citet{Gruppioni:2013jna} are defined by fitting to modified Schecter functions, with parameters that are defined piecewise on $z$. These piecewise fits have breaks at the redshifts $z=0.5$ and $z=1.1$, so it is at these redshifts that we see peaks in observables like the SFRD.
    }
    \label{fig:lum_func}
\end{figure}

In Fig.~\ref{fig:lum_func}, we show the luminosity density as functions of redshift. In particular, we show the quantity $\int dL \: {dN}/{d\log L} $ for both spiral and starburst galaxies, as well as their sum. Assuming a Salpeter-like IMF, the calibration factor is constant, so this quantity is proportional to the SFRD, as can be seen clearly below in Eq.~\eqref{eq:rsf}. 
We additionally show the luminosity dependence of this integrated quantity by presenting contributions from three luminosity ranges. 

\begin{figure}
    \centering
    \includegraphics[width=\columnwidth]{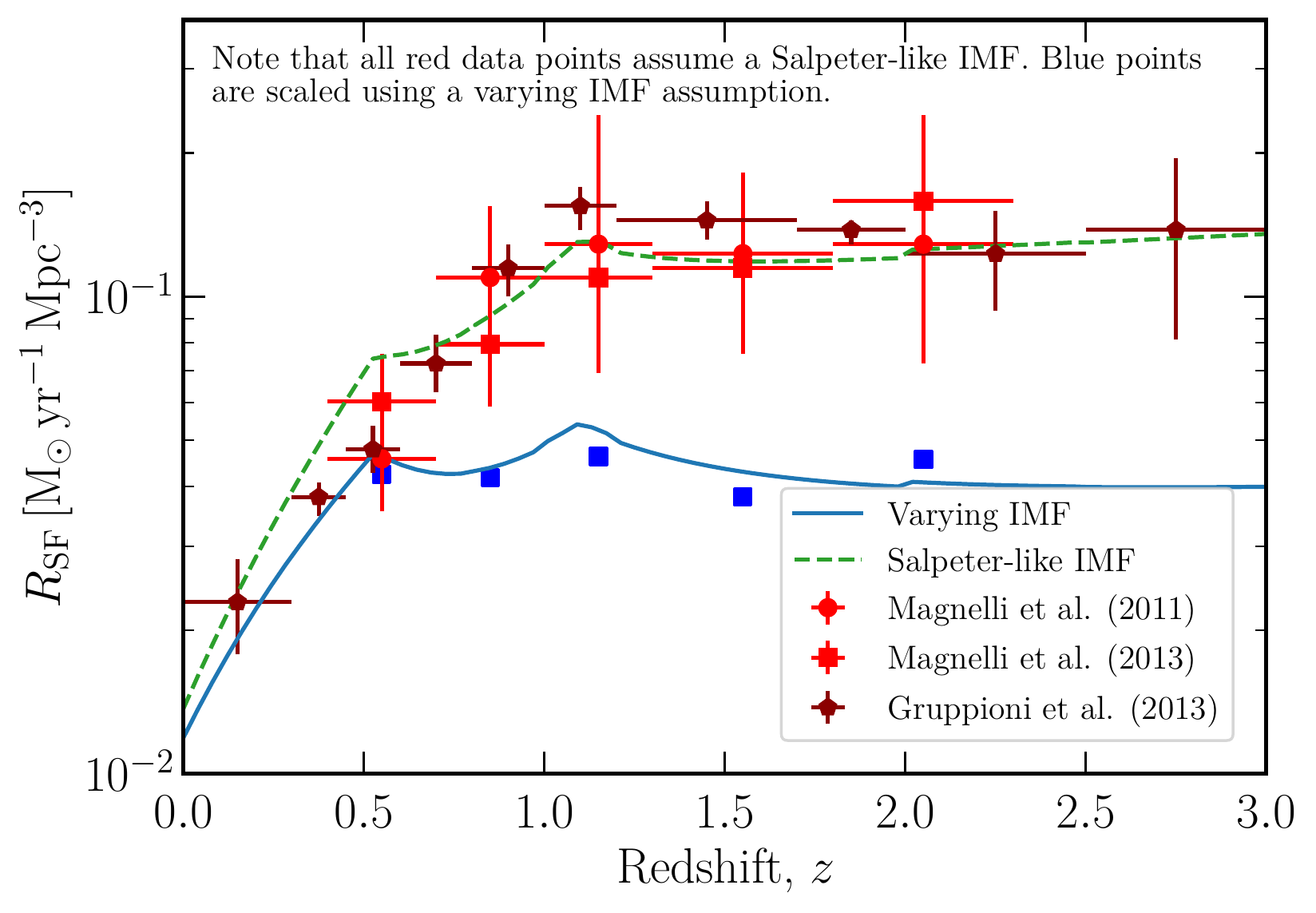}
    \caption{\textbf{Star formation rate density:} We compare the SFRD assuming a Salpeter-like IMF (green dashed curve) and a varying IMF (blue solid curve) to observational data as a function of redshift. The data were chosen to match with the FIR band we explore in this work. To limit clutter in the plot, we chose a representative subset from the catalogue of data~\citep{Madau:2014bja}. Features in the prediction curves at $z=0.5$, $z=1.1$ arise from non-smooth features in the luminosity functions we consider, and features at $z=1.0$, $z=1.2$, and $z=2.0$ arise from a non-smooth division of galaxies with an AGN into spiral and spheroidal sub-categories. While the observed data (red colored points) agrees quite well with the SFRD predicted from a Salpeter-like IMF, the data themselves are derived quantities which assume a Salpeter, Kroupa, or Chabrier IMF. For this reason, it is unsurprising that the observed data do not match the SFRD predicted using the varying IMF, particularly at high redshift. If a varying IMF is assumed when calculating the SFRD from observations, the results are expected to closely follow the predictions we make for a varying IMF. For illustration, we perform a preliminary reanalysis of the \citep{2013A&A...553A.132M} data using the luminosity functions described therein and the varying IMF we use in this work, depicted as blue points. This illustration is meant only to show proof of concept, and a more careful reanalysis should be performed, particularly in order to estimate uncertainties.} 
    \label{fig:SFR}
\end{figure}

From the calibration factor and luminosity function, we can calculate the SFRD for a given galaxy morphology $g$ by integrating over the luminosity $L$:
\begin{equation}
    R_{\rm SF,\,g} = \int \chi(L) L \frac{dN_g}{d\log L}d\log L\,.
    \label{eq:rsf}
\end{equation}
The total observed SFRD is then given by the sum over galaxy types $\sum_{g}{R_{\rm SF,\,g}}$, where $R_{\rm SF,\,g}$ is the SFRD contribution from galaxies of type g.\footnote{
    For simplicity, for the rest of this paper we drop the subscript g notation. All observable quantities defined below implicitly sum over galaxy type.} 
In Fig.~\ref{fig:SFR}, we compare the SFRD calculated using the varying IMF and Salpeter-like IMF. We also plot existing estimates of the SFRD, with data drawn from \citet{2011A&A...528A..35M,2013A&A...553A.132M,Gruppioni:2013jna}. Note that we only show data that explicitly uses the same FIR range as we consider for our luminosity functions. In principle, there is a wealth of other data from different wavelengths to compare to (e.g. \citet{Fermi-LAT:2018lqt, Madau:2014bja, 2018MNRAS.475.2891D, 2006ApJ...651..142H, 2007MNRAS.379..985F}). We leave a more careful comparison to these data sets to future work.

At low redshift ($z\lesssim0.15$), the SFRD calculated using a varying IMF is slightly lower than (within about 20\% of) the SFRD calculated using a Salpeter-like IMF and is consistent with observations. The primary reason for this behavior is that at low redshifts, the dominant contribution to the luminosity functions comes from spiral galaxies with intrinsically lower galactic luminosity, which favor an IMF power law of $\alpha\approx-2.35$. At higher redshifts, the luminosity is dominated by a smaller density of intrinsically more luminous starburst galaxies, which favor a shallower IMF. Accordingly, we see that at redshifts $z \gtrsim 2$, the SFRD calculated using a varying IMF is up to a factor of three lower than the SFRD calculated from a Salpeter IMF and the reported observational data. However, all of the data plotted in Fig.~\ref{fig:SFR}, which is illustrative of much of the data in the literature (e.g. \citet{Fermi-LAT:2018lqt, Madau:2014bja, 2018MNRAS.475.2891D, 2006ApJ...651..142H, 2007MNRAS.379..985F}), assume an IMF with Salpeter-like behavior at high stellar mass. In particular, in each of the three observed data sets shown here, the SFRD was calculated while assuming either a Chabrier IMF, a Kroupa IMF, or a pure Salpeter IMF. 
It is unsurprising then that the data so closely matches the Salpeter-like IMF, while disagreeing with results obtained using any other IMF. We therefore expect that reanalyzing the SFR observations with a varying IMF would result in an SFRD that is lower than the existing data and that matches our predictions. We leave this analysis to future work. Since the effect of a varying IMF is signficant, we also note that direct observations of the SFR at high redshifts would provide a useful probe of the IMF variability as a function of galaxy type.

\subsection{Extragalactic Background Light}

In addition to measuring the SFRD directly, we can also look at the extragalactic background light (EBL) to potentially probe the IMF. The EBL is the integrated light from all sources in a particular direction. In particular, it includes light from all galaxies, even those too faint to resolve, and therefore provides an accurate measure of the luminosity function. In fact, the total EBL can be calculated as
\begin{equation}
    L_{\rm EBL,total} = \int L\frac{d n_g}{d L} dL = \int \frac{d n_g}{d\log L} dL\,.
\end{equation}
That is, the total EBL is simply an integral of the luminosity function over galaxy luminosities. Starting from the luminosity functions then, we can calculate the total EBL without introducing a dependence on the IMF, and so it is impossible to probe the IMF from the total EBL.

However, while the total EBL may be independent of IMF, the IMF affects the distribution of stellar masses. Because stars of different masses have different temperatures, and therefore different spectra, it is possible that the EBL spectrum may provide a way to probe the IMF. With this in mind, we followed the procedure outlined in \citet{Razzaque2009} to calculate an estimate of the EBL flux in the wavelength range 0.1 to 100 $\mathrm{\mu m}$. Specifically, we calculate the spectrum of a star of given mass $M$ with the corresponding effective temperature $T(M)$ as a blackbody spectrum $I_{\nu, BB}(T(M))$.  We additionally denote the number of stars of a given mass inside of a galaxy as $\mathcal{N}(M,L)$. This number therefore intrinsically depends on the IMF. The EBL at a given frequency can then be calculated as 
\begin{equation}
    I_{{\rm EBL}, \nu} =\int \int I_{\nu,BB}(T(M)) \mathcal{N}(M,L) dM \frac{d N}{d\log L} dL\,.
\end{equation}  

\begin{figure}
    \centering
    \includegraphics[width=\columnwidth]{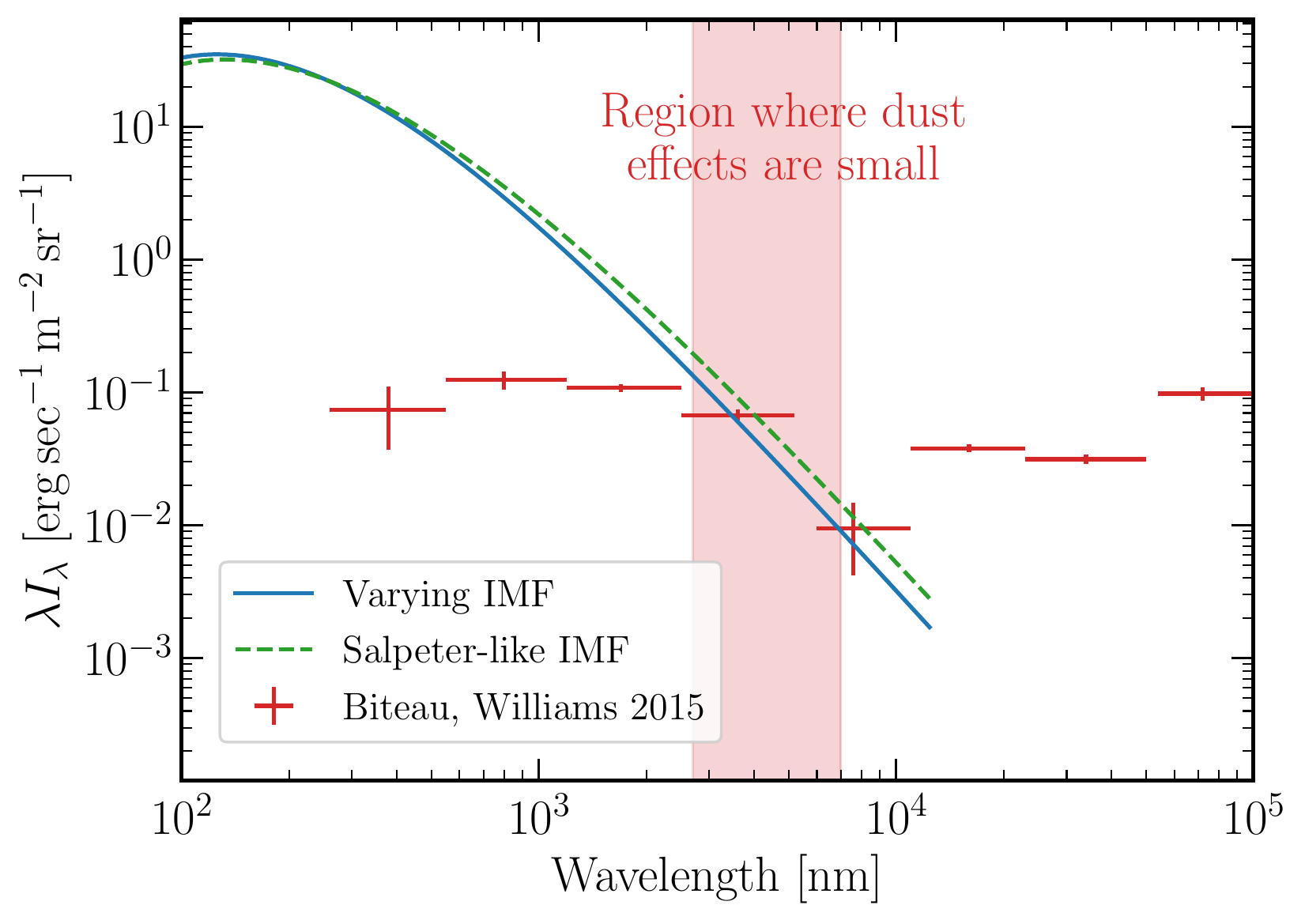}
    \caption{\textbf{Extragalactic Background Light Spectrum:} Estimates of the EBL spectra assuming a varying IMF (blue solid curve), and a universal Salpeter-like IMF (green dashed curve). These are compared to data from \citet{BW2015}, showing the best fit values of the EBL calculated from various observed measurements of upper and lower bounds on the EBL spectrum. The estimates we show do not include processing from dust, so we generally expect the estimates to overpredict the intensity at short wavelengths, but begin to match at longer wavelengths. At the longest wavelengths, radiation from dust, rather than from stellar sources, will dominate the observed spectrum,
    and our estimates will underpredict the data. This behavior is seen in our predictions which, in the wavelength range $3-7 \,\mu$m which is least affected by dust, agrees quite well with data, regardless of IMF considered.}
    \label{fig:ebl}
\end{figure}

Although this technique can give an estimate of the EBL emitted from galaxies, it does not take into account dust, and because of this offers only limited insight into the observed EBL. Absorption of starlight by dust causes the short-wavelength end of the spectrum to be reduced, while re-emission by that dust causes the long-wavelength end to be increased. 
While this generic picture is true in all dust models, exactly how dust affects the shape of the spectrum depends heavily on the dust model. 
We leave careful accounting of these dust effects to future work. Instead, we can look at the narrow range of frequencies where the effects from both dust absorption and emission on the EBL are minimized. In particular, we consider the range approximately $4-7 \,\mu$m~\citep{1989ApJ...345..245C, Kennedy_2013}. As can be seen in Fig.~\ref{fig:ebl}, in this wavelength range, the EBL predicted from both a Salpeter-like and varying IMF are quite similar, with the varying IMF prediction being very slightly bluer than the Salpeter-like prediction, and both visually fit existing data equally well. 

\subsection{Core-Collapse Supernova Rate}
\label{sec:CCSNe}

Along with the SFRD and EBL, we can look at the rate at which stars collapse into supernovae as a way to constrain the IMF. Stars which begin their life with a mass between 8 and approximately 125~$\Msun$ are expected to end their life as a core-collapse supernova (CCSN)~\citep{Heger:2002by}. Standard CCSNe occur when their iron or oxygen-neon-magnesium core, produced through nuclear fusion, exceeds the Chandresekhar mass limit and rapidly collapses~\citep{2017RSPTA.37560271C, Janka:2017vcp, Burrows:2020qrp}. 
This facilitates electron capture on nuclei and free protons in the center of the collapsing core, which makes matter more neutron rich and produces a copious number of electron neutrinos~\citep{Bethe:1979zd, Fuller:1981mu}. Once the density reaches the nuclear saturation density (approximately $2.6 \times 10^{14}~\mathrm{g}~\mathrm{cm}^{-3}$~\citep{Bethe:1979zd}) the strong nuclear force stops the collapse of the inner core~\citep{1979NuPhA.324..487B, 2017RSPTA.37560271C, Bethe:1979zd, BranchWheeler}.
The sudden halt causes the core to bounce and launches a shock wave, which carries stellar material away from the core.
The shock stalls after losing energy through the dissociation of heavy nuclei but can be re-energized by neutrinos, which revive the shock leading to a successful explosion. 
This is the so-called neutrino-driven delayed explosion mechanism~\citep{1966ApJ...143..626C, 1985ApJ...295...14B}.

Eventually, the electromagnetic radiation emitted by the material ejected in a successful explosion is observed as a supernova. Depending on the chemical makeup of the ejected material, CCSNe are classified as Type Ib, Ic, or II supernovae~\citep{Turatto2003, Smith:2014txa}. The supernova can leave behind either a neutron star or black hole depending on how much energy is imparted to the shock wave by neutrinos, compared to the gravitational binding energy of the outer layers of the star. Because some of these types of supernovae will be more detectable than others and any stars that collapse directly to a black hole do not produce a supernova, the fraction of core collapses that can be observed through electromagnetic signatures will depend on the fate of the collapsing massive star, and that in turn may depend on the IMF. 


As with the SFR, we will again focus on the core-collapse supernova rate density (CCSNRD), which allows us to focus on only the redshift dependence. On timescales that are long compared to the lifetime of a star massive enough to undergo core collapse ($\lesssim 10$~Myr), we can assume that when one star goes supernova another star of equal mass is formed. As a result, we can calculate the CCSNRD directly from the SFRD of stars with mass greater than approximately $8\,\Msun$. In particular,
\begin{equation}
    R_{{\rm CCSN}}(z) = \int \chi(L) L \frac{dN}{d\log L} \frac{\int_{8 \mathrm{\Msun}}^{M_{\mathrm{max}}} \xi(M) dM}{\int_{M_{\mathrm{min}}}^{M_{\mathrm{max}}} M\xi(M) dM} d\log L.
    \label{eqn:RCC}
\end{equation}
Here, $M_{\mathrm{min}}$ and $M_{\mathrm{max}}$ refer to the minimum and maximum masses stars can take in the IMF we consider. In all of our calculations we use $M_\mathrm{min} = 0.1\,{\Msun}$ and $M_\mathrm{max} = 125\,{\Msun}$, respectively the approximate lowest mass at which stars can fuse hydrogen and a somewhat arbitrary high mass cutoff that does not significantly affect our results. The factor of $M$ in the denominator is necessary because SFRs measure the total mass of stars that form, whereas the core-collapse rate measures the number of supernova events. 

\begin{figure*}
    \centering
    \includegraphics[width=0.49\linewidth]{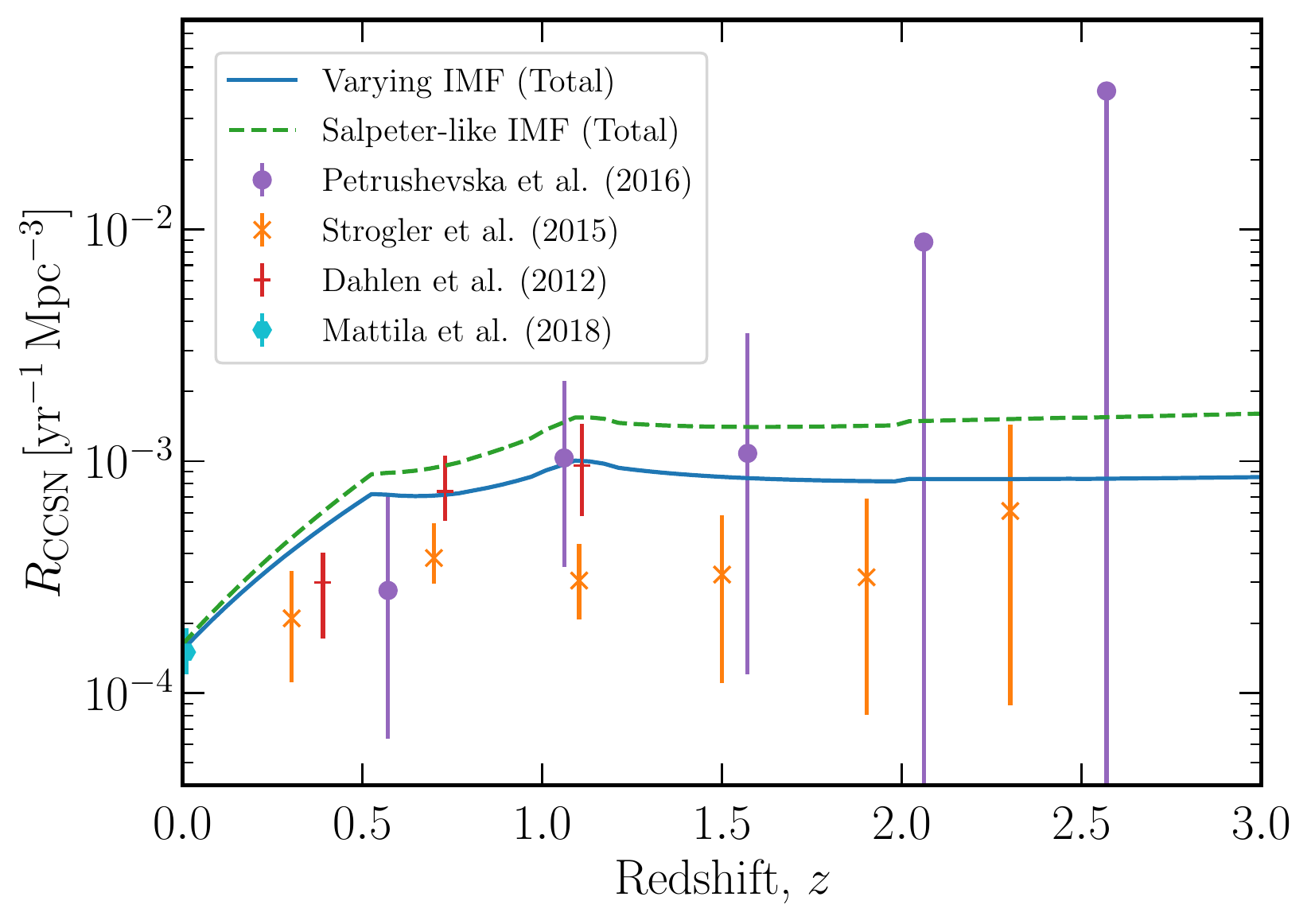}
    \includegraphics[width=0.49\linewidth]{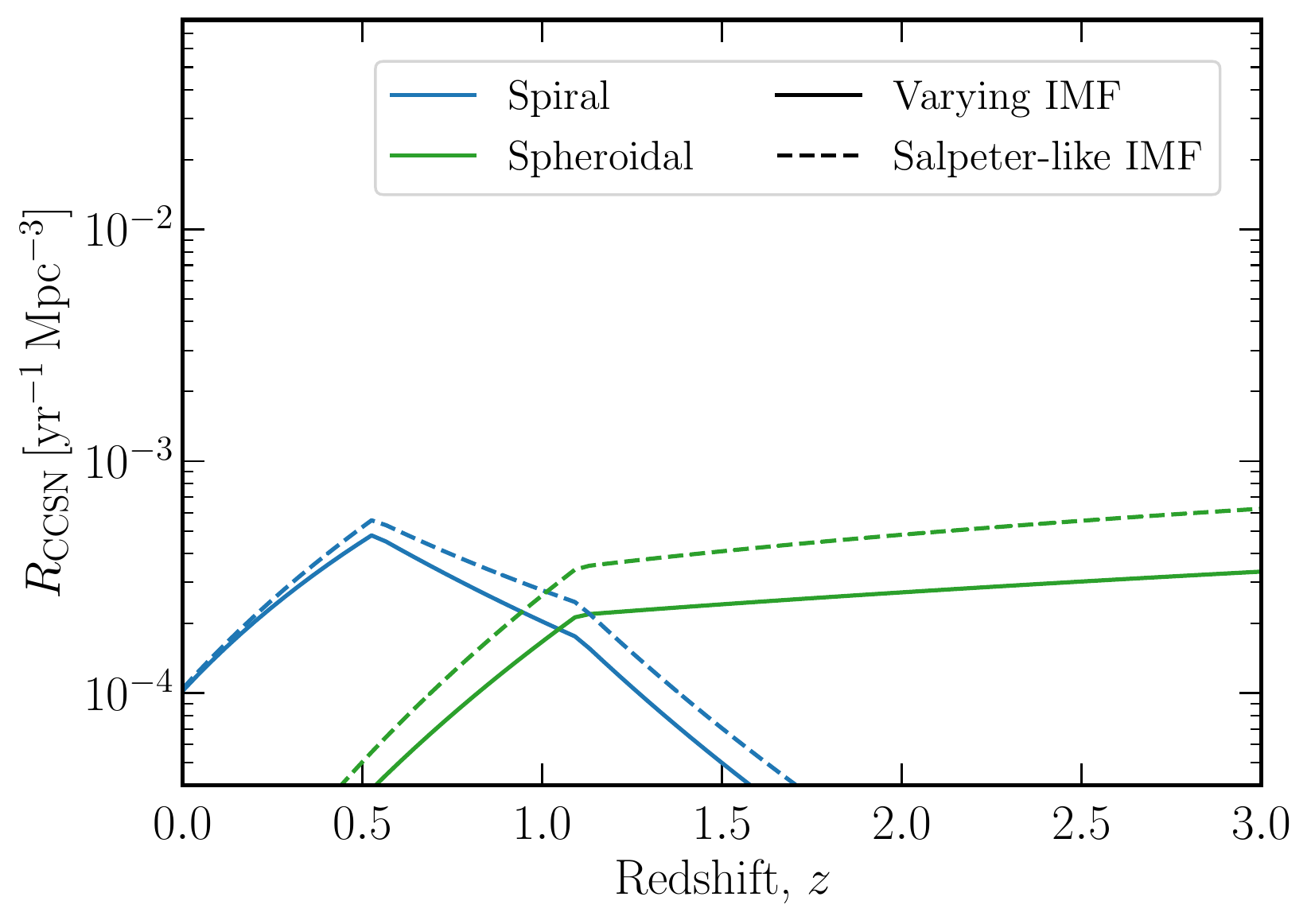}
    \caption{\textbf{Core collapse supernova rate density:} \textit{(Left Panel)} Comparison of the predictions of the rate density of core-collapse supernovae assuming either a Salpeter-like IMF (green dashed curve) or a varying IMF (blue solid curve) with observational data as a function of redshift. See Fig.~\ref{fig:SFR} for discussion of features at $z=0.5$, $z=1.0$, $z=1.1$, $z=1.2$, and $z=2.0$. \textit{(Right Panel)} Rate of supernovae in each of the galaxy morphologies we consider. Here, blue corresponds to spiral galaxies, while green corresponds to spheroidal starburst galaxies. Again, dashed lines correspond to Salpeter-like IMFs, while solid lines correspond to the varying IMF.}
    \label{fig:RCC}
\end{figure*}

In Fig.~\ref{fig:RCC}, we compare the CCSNRD we predict from a varying IMF to that predicted from a Salpeter-like IMF and to observed CCSNRD data from \citet{Petrushevska:2016kie,Strolger:2015kra,Dahlen2012,Mattila2012}. As in the case of SFRs, at low redshift, all three CCSNRD are consistent. At high redshift, the CCSNRD predicted using the varying IMF is slightly lower than that predicted using the Salpeter-like IMF, but the discrepancy between the CCSNRD is significantly smaller than the related difference between SFRs. This improved agreement is to be expected because the SFR's dependence on the IMF partially cancels against the explicit IMF dependence in Eq.~\eqref{eqn:RCC}, as discussed in \citet{Madau:2014bja}. Ultimately, the existing CCSNe data appears to agree with either IMF model equally well. However, as new observatories including, the James Webb Space Telescope~\citep{2019ApJ...874..158R}, the Vera Rubin Observatory (through the LSST survey)~\citep{2019ApJ...873..111I}, Euclid~\citep{laureijs2011euclid}, and the Nancy Grace Roman Telescope~\citep{koekemoer2019ultra} begin to take data, estimates of the high redshift CCSNRD will become more precise, potentially favoring one model over the other. Likewise, gravitational wave detectors may offer another avenue to probe the supernova rate, and consequently the SFR, by measuring the binary black hole merger rate~\citep{Vitale2019}. 

\begin{figure}
    \centering
    \includegraphics[width=\columnwidth]{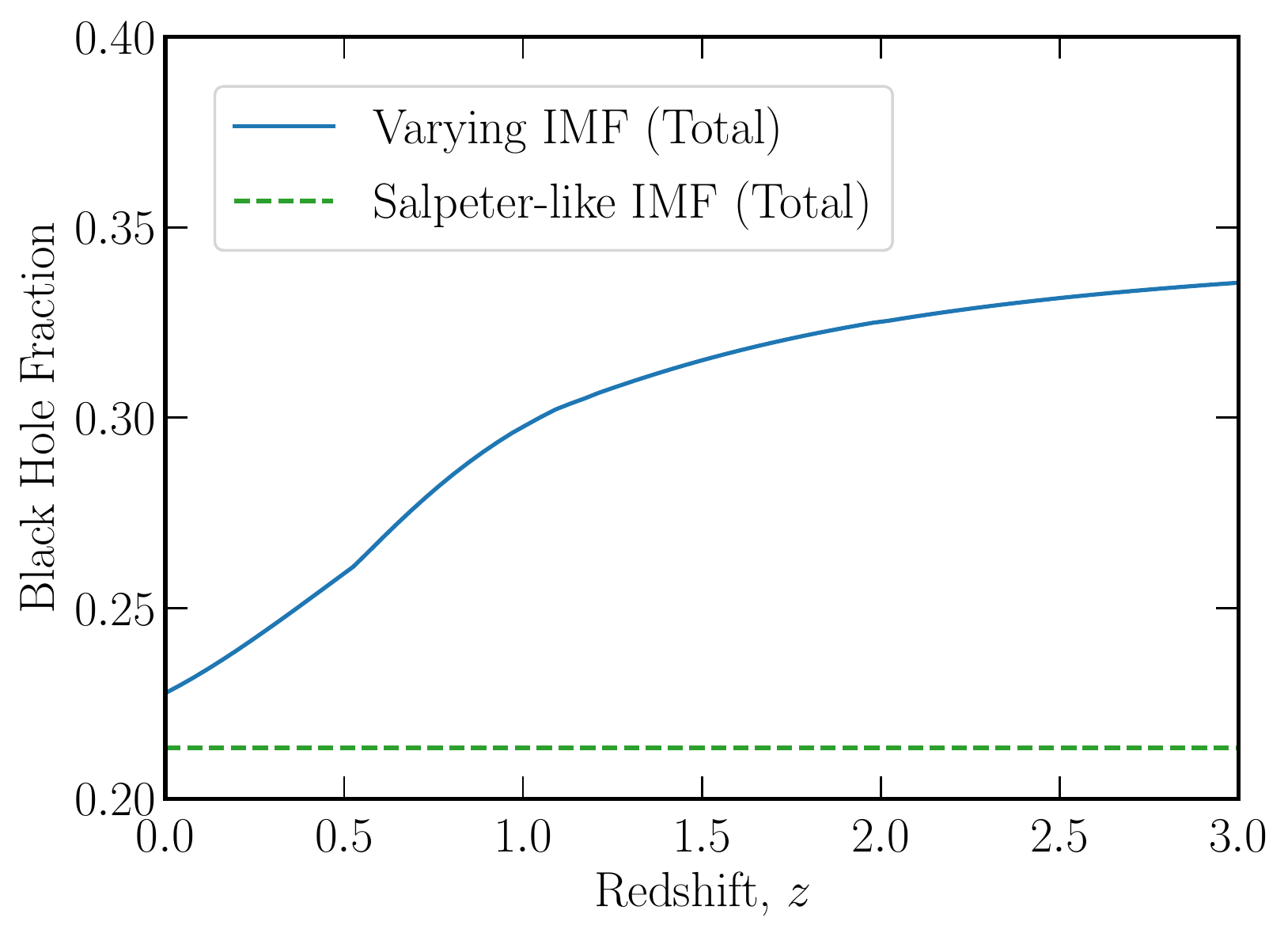}
    \caption{\textbf{Fraction of black hole forming collapses:} The fraction of supernovae that result in black holes, rather than neutron stars. For the Salpeter-like IMF, approximately $21\%$ of stellar collapses  lead to black hole formation (this fraction is assumed to be constant as a function of the redshift for simplicity), while for the varying IMF, this fraction depends on redshift, reaching approximately 35\% at $z=3$.}
    \label{fig:bhfrac}
\end{figure}

We find that the two IMF models produce an $\mathcal{O}(2)$ difference in the total CCSN rate at redshifts greater than approximately 1.5. Therefore, at least a 100\% determination of the SN rate for $z>1.5$ will be required to distinguish these two scenarios. Based on projections of the rates of supernovae expected to be observed by the Roman Telescope~\citep{https://doi.org/10.48550/arxiv.2111.03081}, it is not unreasonable to expect observational uncertainties at that level. Furthermore, a greater difference in the predicted supernova rates under our two models may appear if changing the IMF changes the fraction of stars that collapse into black holes versus those that collapse to neutron stars. All else held equal, a shallower IMF would increase the fraction of stars that evolve to black holes and decrease the fraction that evolve to neutron stars, relative to a steeper IMF. As shown in Fig.~\ref{fig:bhfrac}, this is precisely the situation we expect from the varying versus Salpeter-like IMFs. As a result, we can expect that a varying IMF would lead to fewer obervable CCSNe than a Salpeter-like IMF, particularly at high redshifts, all other factors held equal. 


\subsection{Type Ia Supernova Rate}

As with CCSNe, we can hope to probe properties of the IMF by looking at Type Ia supernovae (SNeIa). Type Ia supernovae occur when mass accretes onto white dwarfs from a giant star, typically when the two form a binary~\citep{Mazzali2007}. As the mass of the white dwarf approaches the Chandrasekhar mass~\citep{Chandrasekhar:1931ih}, temperatures and densities within the core of the star become high enough to initiate nuclear reactions from the abundant carbon and oxygen. These nuclear reactions produce so much energy that the entire star becomes unbound. Nuclear decays within the ejected material can then be observed as a SNeIa. 

For our purposes, SNeIa can be seen as a tracer of SFR like the CCSN rate. In addition, measurements of the SNIa rate do not require one to use a calibration factor and are therefore relatively independent of the IMF. However, since SNeIa are sourced by white dwarfs, which form from stars whose lifetimes are on the order of 1-10 Gyr, they can only occur after enough time has passed for the white dwarf to accrete sufficient mass. Both the lifetime of the progenitor star and the period between the formation of a white dwarf and the occurrence of a SNIa must be accounted for when calculating the SNIa rate. This accounting is done through a delay-time distribution (DTD) $F(\tau)$, which measures the probability that a star formed at some time $t-\tau$ will undergo a supernova at time $t$. In our calculations, we use the simple approximation that $F(\tau) \propto \tau^{-1}$, justified in \citet{Maoz2014}.
The SNIa rate density can therefore be calculated from the convolution
\begin{equation}
    R_{\rm SNIa} = \int_0^t R_{\rm SF}(t-\tau) F(\tau) d\tau\,.
    \label{eqn:RSN1a}
\end{equation}
Here, we convert between redshift and time assuming a flat $\Lambda$CDM Universe that is dominated by $\Lambda$ and matter, and where $\Omega_{m,0} = 0.308$~\citep{Planck2016}. 

\begin{figure}
    \centering
    \includegraphics[width=\columnwidth]{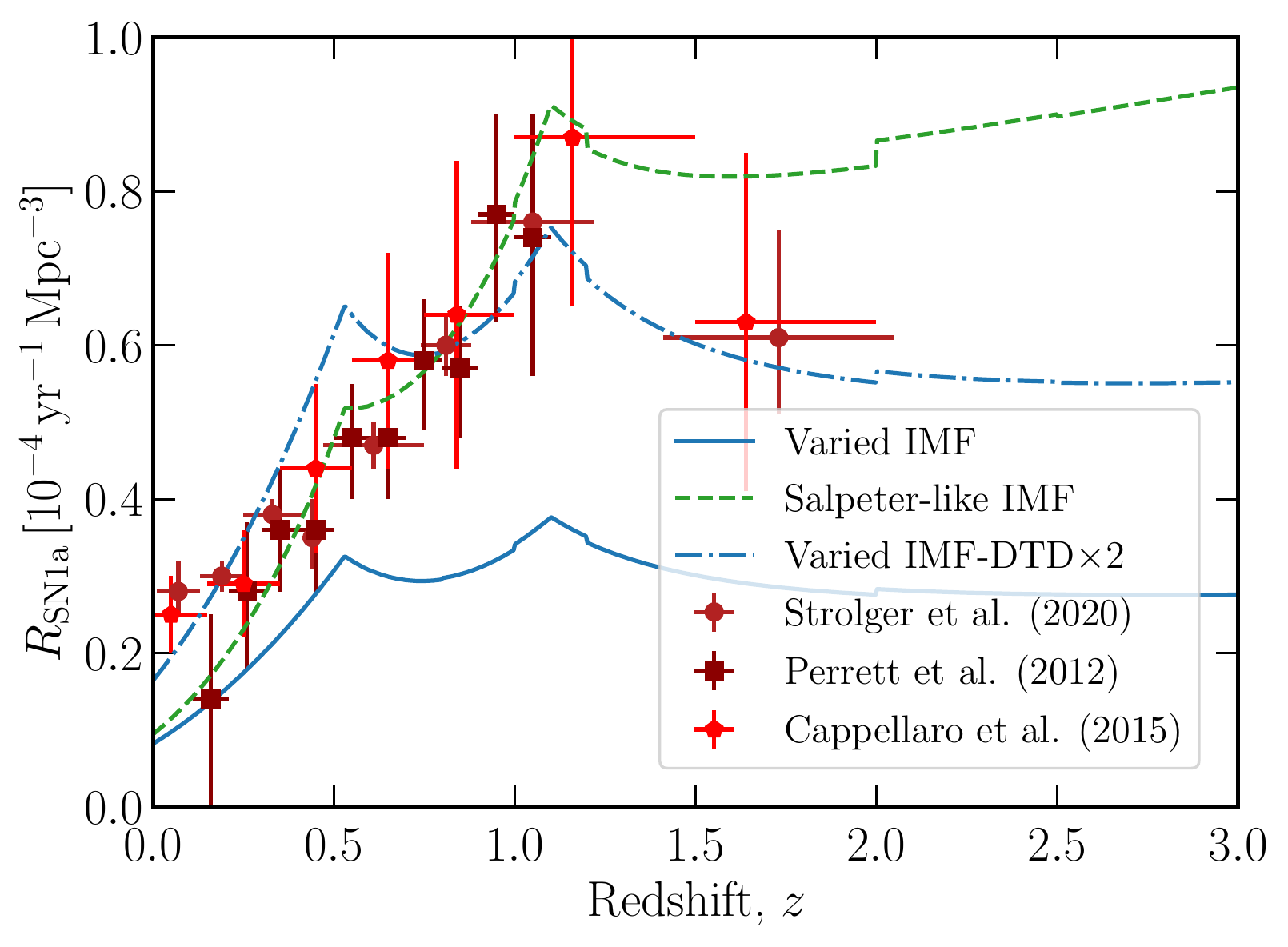}
    \caption{\textbf{Type Ia supernova rate density:} We compare the rate density of type Ia supernovae, assuming either a Salpeter-like IMF (green dashed) or varying IMF as described in the text (blue), to a selection of observed data. See Fig.~\ref{fig:SFR} for discussion of features at $z=0.5$, $z=1.0$, $z=1.1$, $z=1.2$, and $z=2.0$. Predictions of the supernova rate depend on a poorly constrained delay time distribution, and by increasing the overall normalization of this DTD by a factor of 2 (still within $1\sigma$ uncertainties), we can increase the varying IMF result from the blue solid curve to the blue dot-dashed curve.}
    \label{fig:SN1a}
\end{figure}

In Fig.~\ref{fig:SN1a} we compare observed SNIa rates, using data from \citet{Strolger2020, Perrett2012, Cappellaro2015}, with predictions of the supernova rate assuming a Salpeter-like IMF and a varying IMF. To make this comparison, we fixed the normalization of the DTD to $10^{-3} \, \Msun^{-1}$ in order to match the DTD presented in \citet{Maoz2014}.\footnote{However, as noted in \citet{Maoz2014}, measurements of the DTD in different environments (e.g. dwarf galaxies, galaxies, and galaxy clusters) can vary by an order of magnitude.} Using this normalization, the Salpeter-like IMF is largely consistent with the data, while the varying IMF leads to a SNIa rate that is consistently smaller than the observed SNIa rate. However, if we allow the normalization of the DTD to vary, then increasing it by a factor of 2 produces much closer agreement between the varying IMF and observed SNIa rates. This change to the normalization is 
still within the $1\sigma$ confidence range for the DTD. In other words, based on current observations and their uncertainties, the varying IMF cannot be ruled out, and more precise measurements of the DTD would be necessary to place meaningful constraints on either IMF model.

\subsection{Diffuse Supernova Neutrino Background}
\label{sec:DSNB}

Finally, in addition to directly observing of supernova rates, we can look to the neutrinos supernovae produce in order to estimate their rate, and therefore potentially probe the IMF. Despite being relatively rare in any individual galaxy, supernovae are quite common throughout the Universe. Combining this with the fact that a single CCSN produces an immense number of neutrinos (approximately $10^{58}$~\citep{Burrows:2020qrp, Mirizzi:2015eza}) leads to the emergence of a background of neutrinos that is isotropic  and nearly constant in time. This neutrino flux is commonly named the diffuse supernova neutrino background (DSNB)~\citep{1984NYASA.422..319B, Krauss:1983zn, Wilson:1986ha, Beacom2010, Lunardini:2010ab, Mirizzi:2015eza}. 

Only in recent years have experiments begun to approach the sensitivity necessary to directly observe the DSNB~\citep{SuperKamiokande2003, Bays:2011si, Zhang:2013tua, KamLAND:2021gvi, Super-Kamiokande:2021jaq,Li:2022myd}. While no signal has yet been detected, the enrichment of Super-Kamiokande (SK) with gadolinium~\citep{Beacom:2003nk, Horiuchi:2008jz} and the future proposed and planned experiments such as Hyper-Kamiokande (HK), JUNO, Jinping, and THEIA~\citep{JUNO:2015zny, Jinping:2016iiq, Hyper-Kamiokande:2018ofw, Sawatzki:2020mpb, Li:2022myd} are expected to have enough sensitivity to make a first detection in the coming years. Once observed, the DSNB will provide a test of astrophysical observables~\citep{Lunardini:2009ya, Keehn:2010pn, Nakazato:2013maa, Nakazato:2015rya, Priya:2017bmm, Moller:2018kpn, Kresse:2020nto, Singh:2020tmt, Horiuchi:2020jnc, Libanov:2022yta}, neutrino flavor physics~\citep{Lunardini:2012ne, Tabrizi:2020vmo, Suliga:2021hek}, and physics beyond the Standard Model~\citep{Ando:2003ie, Fogli:2004gy, Goldberg:2005yw, Baker:2006gm, Farzan:2014gza, Jeong:2018yts, Creque-Sarbinowski:2020qhz, DeGouvea2020, Das2021, Suliga:2021hek, deGouvea:2022dtw}.

\subsubsection{Theoretical Models}
\label{sec:Neutrino Spectra}

The calculation of the DSNB flux requires two components: the rate of supernovae as a function of their progenitor masses and the time-integrated neutrino energy spectra associated with each supernova. The former can be calculated as described in Section~\ref{sec:CCSNe}, while a calculation of the latter is sketched here. Following \citet{Moller:2018kpn,Ashida:2022nnv}, we consider three possible outcomes of supernovae, depending on the mass of their stellar progenitor. Stars can either evolve into black holes or low/high mass neutron stars. A characteristic neutrino spectrum is then associated with each type of explosion.  More details on these spectra can be found in Appendix~\ref{app:neutrino}.

As discussed in Appendix~\ref{app:neutrino}, the characteristic neutrino spectra associated with the different supernovae outcomes can be significantly different.
The DSNB signal is therefore also affected by the fraction of black hole forming collapses~\citep{Lunardini:2009ya}. Unfortunately, the fraction of stellar collapses leading to black hole formation is unknown. Recent theoretical work and observational surveys indicate that this fraction could be approximately $10\,$--$\,40\%$ of all CCSNe~\citep{Byrne:2022oik,Neustadt:2021jjt,Kochanek:2008mp, Lien:2010yb, Gerke:2014ooa,Horiuchi:2014ska, Sukhbold:2015wba, Ertl:2015rga, Adams:2016ffj, Adams:2016hit, Davies:2020iom}. 
For the Salpeter-like IMF, our DSNB modeling follows one of the scenarios considered in \citet{Moller:2018kpn} where the fraction of progenitors evolving into black holes is set to $21 \%$ and for simplicity it is assumed to be constant as a function of redshift out to at least $z=5$. To make this assumption, we implicitly assume that both the stellar masses that evolve to black holes and the IMF do not change over cosmological history. However, when we consider the varying IMF, we cannot assume that the IMF is constant as a function of redshift; in fact, because it depends explicitly on the star formation rate, which evolves over cosmic history, the varying IMF cannot remain constant. 
As a result, we expect to see a substantially larger fraction of stars evolve to black holes at high redshifts in the non-universal case than in the Salpeter-like case, as seen in Fig.~\ref{fig:bhfrac}.\footnote{
    It should be noted that this fraction is calculated assuming that the stellar mass is the only factor which controls whether a supernova leads to a black hole formation or not (see Appendix~\ref{app:neutrino}.). It is therefore constant for the universal Salpeter-like IMF. Other stellar properties, such as metallicity, may lead to changes in the fraction of supernovae that lead to black hole formation.}
In addition, because our main goal is to investigate the impact of the variable IMF on the DSNB, we do not include neutrino oscillations in our modeling, as they would change the DSNB in the same way in both cases, i.e., Salpeter-like and variable IMF. 

Making use of the supernova rate and the neutrino spectrum, we can calculate the DSNB flux $\Phi$ as a function of neutrino energy $E$. Specifically, 
\begin{equation}
\label{eqn:DSNBsimple}
\begin{split}
\Phi(E) = c &\int_{0}^{z_{\max}} \frac{\mathrm{d} z}{H(z)} \\
&\times\int \chi(L) L \frac{dN}{d\log L} \left[\int_{8M_{\odot}}^{M_{\max}}\frac{\mathrm{d}n}{\mathrm{d}E^\prime} \frac{\xi(M) dM}{\overline{M}}\right] d\log L \,,
\end{split}
\end{equation}
where $\frac{dn}{dE}$\footnote{
    Note that although Eq.~\eqref{eqn:DSNBsimple} is true for all flavors, as mentioned in Appendix~\ref{app:neutrino} we will focus solely on electron-antineutrinos since they have the best detection prospects.
}
is the neutrino spectrum, $E^\prime = E(1+z)$ is the source energy necessary at a redshift of $z$ to be observed with energy $E$, and $\overline{M} = \int_{M_{\min}}^{M_{\max}} M\xi(M)dM$ is the average mass of newly formed stars. The integral $\int \chi(L) L \frac{dN}{d\log L} d \log L$ is equivalent to $R_{\rm SF}$ if the term in brackets does not depend on luminosity. While this condition is met for the Salpeter-like IMF, when we consider a varying IMF, the IMF depends on luminosity, so the factor in brackets picks up a dependence on luminosity. In addition, $H(z) = H_0(\Omega_{m,0}(1+z)^3 + (1-\Omega_{m,0}))^{1/2}$ is the Hubble expansion parameter at $z$, where we assume a flat $\Lambda$CDM universe dominated by matter and $\Lambda$, with $\Omega_{m,0} = 0.308$ and $H_0= 67.8 \, \mathrm{km\, s^{-1}\, Mpc^{-1}}$~\citep{Planck2016}.

\begin{figure*}
    \centering
    \includegraphics[width=0.49\linewidth]{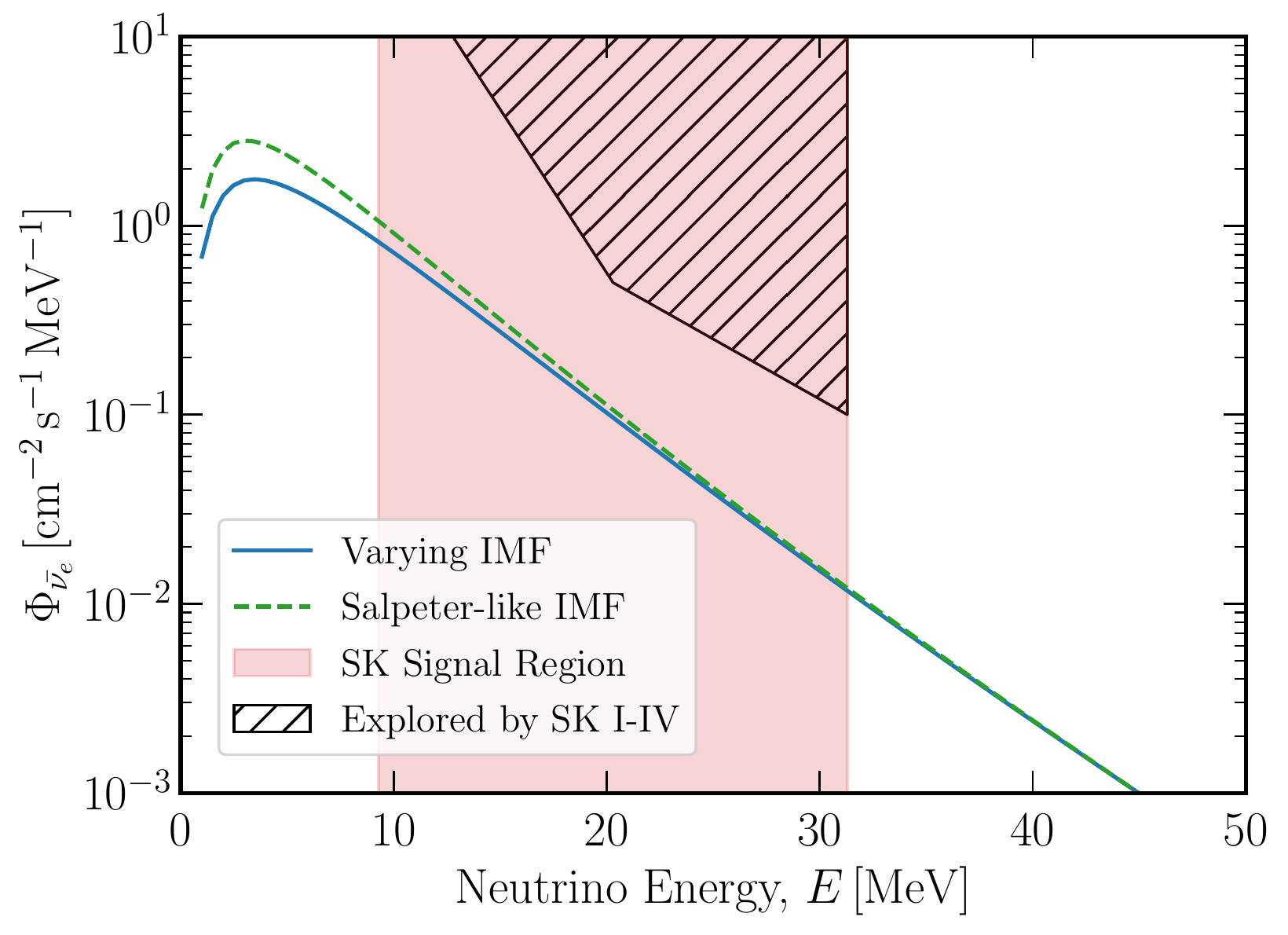}
    \includegraphics[width=0.49\linewidth]{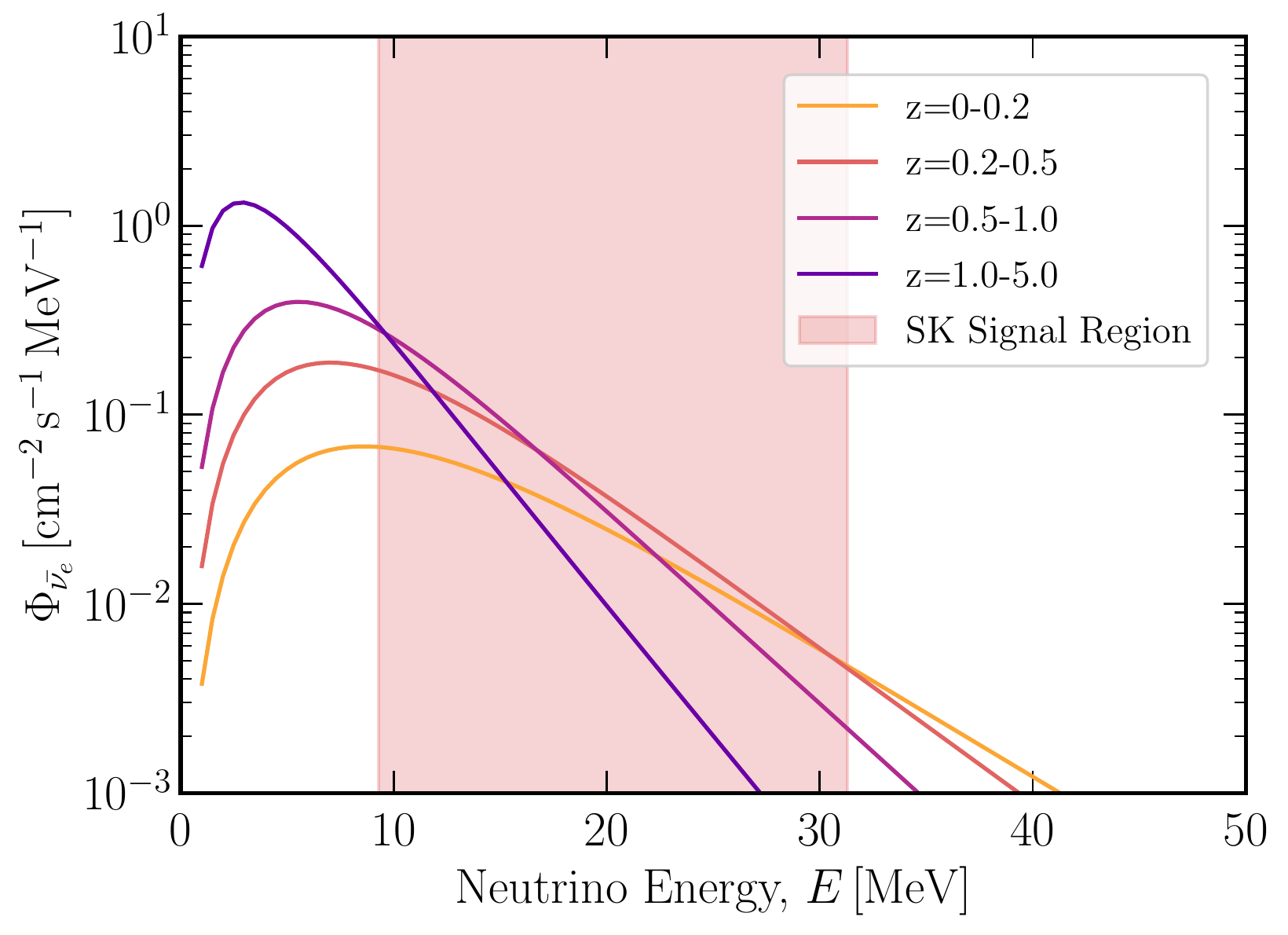}
    \caption{\textbf{Diffuse supernova neutrino background flux:} \textit{(Left Panel)} We compare the predicted DSNB $\bar\nu_e$ flux, assuming either a Salpeter-like (green dashed) or varying IMF (blue solid) with the region probed by SK (hatched) \citep{Super-Kamiokande:2021jaq}. The red shaded region indicates the range of energies that could be observable by SK with the addition of gadolinium. The different IMFs might have an impact on the observed DSNB flux only at low energies, and it would be difficult to distinguish this difference against background neutrino sources. \textit{(Right Panel)} We show the DSNB broken down into contributions from different redshift bins for the varying IMF. Because the emitted neutrinos are redshifted, distant supernovae only contribute significantly at low energies.}
    \label{fig:DSNB}
\end{figure*}

Using the CCSNRD calculated from both a Salpeter-like and varying IMF we find the DSNB fluxes shown in Fig.~\ref{fig:DSNB}. We emphasize here that we have only considered a single benchmark for each IMF model, but in reality there are a number of other uncertainties which would create a band of possible DSNB realizations. We do not directly consider these uncertainties in the modeling of the DSNB (i.e., in Fig.~\ref{fig:Rate}), although below we account for a systematic uncertainty when evaluating the discriminability of the two IMF models. For reference, we also show a hatched region which shows the approximate parameter space currently ruled out by SK observations, as well as the primary signal region for SK and future detectors~\citep{elhedri:hal-03373391}. Perhaps the most notable result is the similarity between the predictions from using the Salpeter-like and varying IMFs, especially at energies greater than approximately 20\,MeV. While this will likely make it difficult to distinguish between the two scenarios once the DSNB is detected, it clearly demonstrates that the DSNB is robust to significant changes to the IMF. 

At lower energies (below approximately $20$ MeV), we predict the DSNB flux to be slightly lower when assuming a varying IMF than when assuming a Salpeter-like IMF. While the IMFs we consider give similar predictions for the CCSNRD at low redshift, they begin to disagree at higher redshifts. The difference in the DSNB flux that we see here ultimately arises from this difference in supernova rates. This is only noticeable at low neutrino energies because distant supernovae only contribute significantly to the low energy spectrum. At higher energies, because of the redshifting of the neutrinos as they propagate, the source energy $E^\prime = E(1+z)$ can be significantly higher for distant supernovae than for nearby ones. As a result, the contribution to the DSNB flux from distant supernovae comes from a higher energy portion of the spectrum, which is exponentially suppressed. This effect is illustrated in the right panel of Fig.~\ref{fig:DSNB} where we show the contribution to the DSNB in different redshift bins. Therefore, while a varying IMF is likely indistinguishable from a Salpeter-like IMF at high energies, sufficiently precise measurements of lower energy neutrinos may allow some degree of distinction between the two models.

Varying the IMF is not the only source of uncertainty expected to appear in the low-energy DSNB region (around $20$~MeV). Variations to the star formation histories, for example, could yield comparable differences between DSNB predictions~\citep{Singh:2020tmt, Kresse:2020nto}, which will be degenerate with differences due to variations of the IMF. Furthermore, additional uncertainties in the DSNB flux which may appear, independent of the choice of IMF, include the unknown fraction of high-mass stars that evolve to black holes and the neutrino spectra emitted during this evolution~\citep{Lunardini:2009ya,Horiuchi:2017qja, Kresse:2020nto}, the precise normalization of the CCSN rate~\citep{Horiuchi:2011zz,Mathews:2014qba}, the evolution of neutrino flavors in the dense medium encountered during supernovae~\citep{Duan:2010bg, Chakraborty:2016yeg, Tamborra:2020cul}, and any possible stellar binary interactions~\citep{Horiuchi:2020jnc}. To partially account for these uncertainties in the following subsection~\ref{Sec:DSNB_IN_DET}, we include a systematic uncertainty of $50\%$ in our calculations of the discriminating power of the detectors to the Varying IMF.

\subsubsection{Expected Sensitivities of Super-Kamiokande and Hyper-Kamiokande}

\label{Sec:DSNB_IN_DET}

\begin{figure}
    \centering
    \includegraphics[width=\columnwidth]{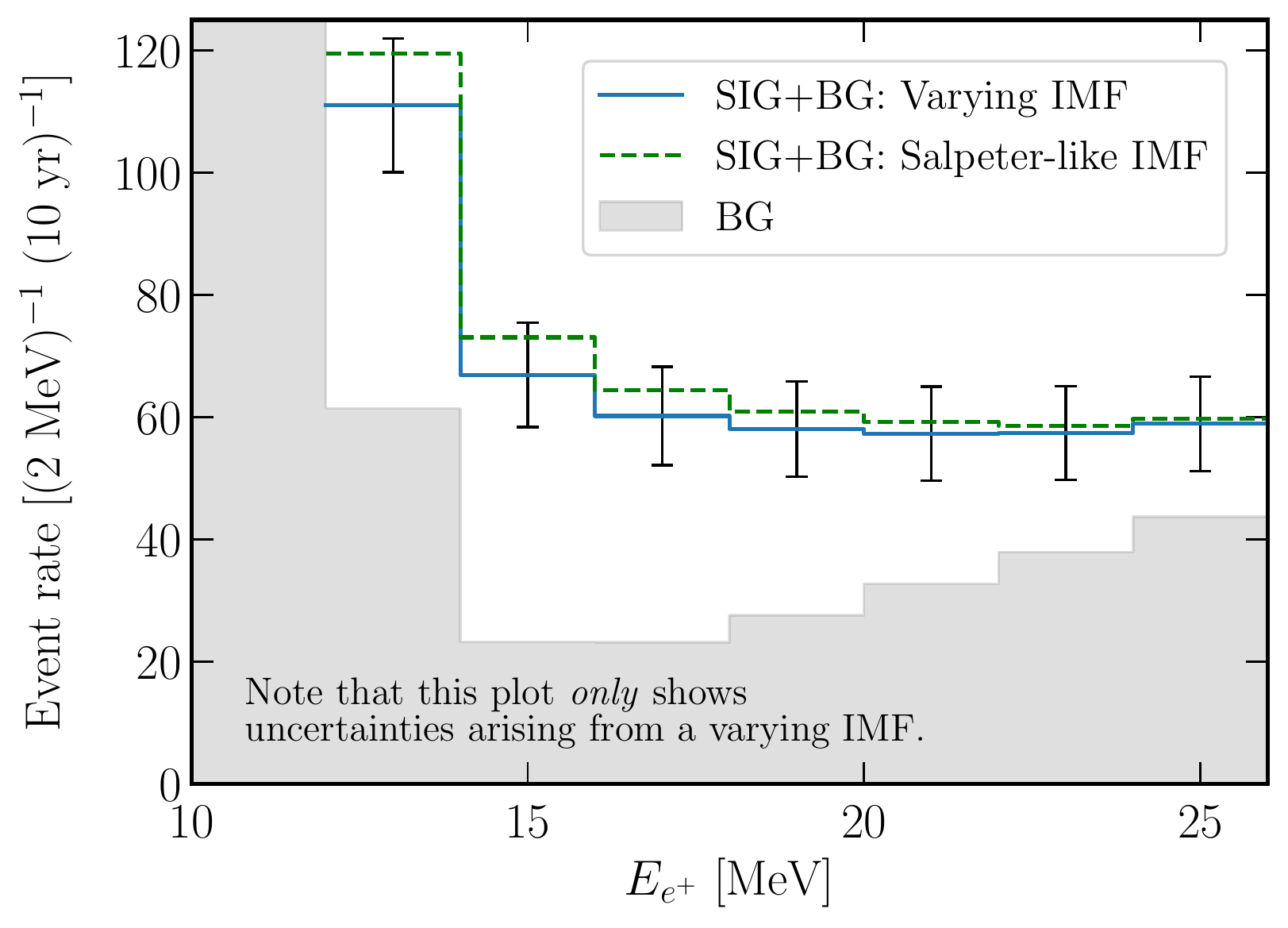}
    \caption{
    \textbf{Diffuse supernova neutrino background rate:} $\bar\nu_e$ DSNB event rates in HK enriched with gadolinium detectors for 10 yrs of data taking. The sum of the $\bar\nu_e$ DSNB event rate plus background rate for the Varying IMF (Salpeter-like) is plotted with solid blue (dashed green) line. The backgrounds rates are depicted as grey regions, and the error bars reflect the $\pm 1 \sigma$ statistic uncertainties. As discussed in the main text, while SK cannot distinguish the two investigated IMF scenarios, HK might present a low significance hint towards a particular scenario. Note that we do not show other astrophysical uncertainties (e.g., errors on the SFR and uncertainty in the neutrino flux modeling), but partially account for this with a systematic uncertainty (see main text for details).}
    \label{fig:Rate}
\end{figure}

Although no detection of the DSNB has yet been made, the strongest constraints come from the SK experiment~\citep{Bays:2011si, Super-Kamiokande:2021jaq}. In 2019, upgrades to SK began which allowed for the introduction of gadolinium into the SK tank by 2021. The gadolinium doping will make the detection of electron antineutrinos significantly easier, which will subsequently improve our ability to detect neutrinos from the DSNB~\citep{Beacom:2003nk}. As a result, it is expected that a positive measurement of the DSNB will be observed in the near future~\citep{Li:2022myd}. 

Figure~\ref{fig:Rate} shows the predicted accumulated DSNB flux after 10 years of operation of HK (3740 kton~yr exposure) compared to its respective neutrino backgrounds, where
 we assume a concentration of 0.1\%  GdCl${}_3$ in water.
The blue solid (green dashed) line depicts the combined neutrino flux from both the DSNB and background sources, where we calculate the DSNB flux using a varying IMF (Salpeter-like IMF).
Sources of background neutrinos that we consider include atmospheric charged-current events, invisible muons, $^9$Li spallation, and reactor antineutrinos~\citep{Hyper-Kamiokande:2018ofw,Super-Kamiokande:2021jaq}. 

We can use a simple $\Delta \chi^2$ Pearson test to estimate the detection prospects of distinguishing between the varying IMF and Salpeter-like IMF. 
In both SK (225 kton yr exposure with 0.1\%  GdCl${}_3$) and HK, the two IMFs are not distinguishable at the $3\sigma$ level, even after 10 years of data collection. Furthermore, in SK, the two models remain indistinguishable at the $1\sigma$ level.
However, in HK, the varying and Salpeter-like IMFs are distinguishable at the $1.3\sigma$ level. This marks an upper bound to the distinguishability, as we have not introduced other sources of uncertainty. Including a systematic uncertainty for the background of 20\% and for the DSNB models of 50\% reduces the difference between the two models to $1.1\sigma$ (note that these additional systematics are not shown in Fig.~\ref{fig:Rate}).
Once SK or HK detects the DSNB, it may be possible to generate an improved statistic, using an unbinned maximum likelihood ratio test with a parameterized family of models that include the Salpeter-like and varying IMFs, which could allow for a better ability to distinguish between these two models, assuming all other uncertainties become negligible in comparison.

\section{Conclusion and Outlook}
\label{sec:conclusion}

In this paper, we explored whether one can find evidence for a non-universal IMF in five astrophysical observables that arise from integrating over cosmological scales. Throughout, we have seen that these observables show small but non-zero differences when assuming the non-universal IMF (described in Sec.~\ref{sec:varIMF}) versus a universal Salpeter-like IMF. The  differences are typically too small to distinguish in currently existing data, but as detectors improve, there are a variety of signals which may offer practical ways to study the IMF. In particular, we find that studying the star formation rate (SFR) and the core-collapse rate at high redshifts ($z\gtrsim0.5$) may offer the greatest distinguishability between the different IMF models.

At redshifts greater than approximately 0.5, the SFR predicted assuming a varying IMF is lower than the corresponding prediction from a  Salpeter-like IMF by up to a factor of $\sim3$ (see Fig.~\ref{fig:SFR}). Because observing the SFR requires a calculation that depends strongly on the assumed IMF, the predictions, though significantly different, can both still be consistent with current observations. However, due to the SFR's prominent role in, for example, modeling star formation histories and cosmological simulations, it would be interesting to see whether an indirect test of the SFR could favor one IMF model over the other. For example, the binary black hole (BBH) merger rate~\citep{Fishbach:2018edt,vanSon:2021zpk} acts as an independent probe of the SFR. LIGO and Virgo, at design sensitivity, are expected to observe BBH mergers up to redshifts of around one. On the other hand, 3rd generation telescopes may reach $z\sim15$ and thus independently measure the star formation rate density (SFRD) to a few percent according to \citet{Vitale2019}. We leave a more careful examination of this method to future work.

Similarly to the SFR, above $z\gtrsim 0.5$ the core-collapse rate can differ by a factor of $\sim2$ between the two IMF models (see Fig.~\ref{fig:RCC}). Current observations of core-collapse rates are still too poor at high redshifts to distinguish between these models. However, next-generation telescopes such as James Webb Space Telescope~\citep{2019ApJ...874..158R}, Roman Space Telescope~\citep{koekemoer2019ultra}, Vera Rubin Telescope (through the LSST)~\citep{2019ApJ...873..111I}, and EUCLID~\citep{laureijs2011euclid},  may provide significantly better rate estimates at these high redshifts. An additional subtlety here is that a varying IMF will produce a larger fraction of black hole collapses (see Fig.~\ref{fig:bhfrac}) which may be less luminous. A careful treatment of the observational efficiency of any core-collapse rate measurement is therefore required to distinguish the two IMF models.

The stellar IMF plays a fundamental role throughout astrophysics and many unsolved questions require an accurate model of the IMF to address. 
It is therefore vital that new methods are developed to decipher its dependence on the local environment. Here we have examined a few indirect methods. Future work should more carefully examine the most promising of these scenarios with a meticulous treatment of their associated uncertainties, including studying whether the impact of different IMFs can be distinguished from potentially degenerate or partially degenerate sources of uncertainty. Moreover, combining the different probes presented here may lead to significantly tighter constraints on possible IMF variations. Through this, we hope that indirect probes may provide additional evidence towards distinguishing a varying IMF.

\section*{Acknowledgments}
We thank Andrew Hopkins for a careful reading of the manuscript and helpful discussions. T.E. and K.F. acknowledge support by the Vetenskapsr{\aa}det (Swedish Research Council) through contract No.  638-2013-8993 and the Oskar Klein Centre for Cosmoparticle Physics at Stockholm University. T.E. was supported by the NWO through the VIDI research program ``Probing the Genesis of Dark Matter'' (680-47-5).
J.Z. and K.F. are grateful for support from the Jeff \& Gail Kodosky Endowed Chair held by K. F. at the University of Texas.  J.Z. and K.F. acknowledge funding from the U.S. Department of Energy, Office of Science, Office of High Energy Physics program under Award Number DE-SC0022021.
A.M.S. acknowledges the support form the US National Science Foundation (Grant No. PHY-2020275). In Copenhagen, this work was supported by the Villum Foundation (Projects Nos. 13164 and 37358), the Danmarks Frie Forskningsfonds (Project No. 8049-00038B), the Deutsche Forschungsgemeinschaft through Sonder- forschungbereich SFB 1258 “Neutrinos and Dark Matter in Astro- and Particle Physics” (NDM).
The work of SH is supported by the US Department of Energy under the award number DE-SC0020262, NSF Grant numbers AST-1908960 and PHY-1914409, and JSPS KAKENHI Grant Number JP22K03630. 
SA was supported by MEXT KAKENHI Grant Numbers,  JP20H05850 and JP20H05861.
This work was supported by World Premier International Research Center Initiative (WPI Initiative), MEXT, Japan. 
A.M.S. and S.H. would like to thank Kavli Institute for Theoretical Physics for the hospitality during this work. This research was supported in part by the National Science Foundation under Grant No. NSF PHY-1748958.

\bibliographystyle{mnras}
\bibliography{DSNB}

\begin{thebibliography}{}
\makeatletter
\relax
\def\mn@urlcharsother{\let\do\@makeother \do\$\do\&\do\#\do\^\do\_\do\%\do\~}
\def\mn@doi{\begingroup\mn@urlcharsother \@ifnextchar [ {\mn@doi@}
  {\mn@doi@[]}}
\def\mn@doi@[#1]#2{\def\@tempa{#1}\ifx\@tempa\@empty \href
  {http://dx.doi.org/#2} {doi:#2}\else \href {http://dx.doi.org/#2} {#1}\fi
  \endgroup}
\def\mn@eprint#1#2{\mn@eprint@#1:#2::\@nil}
\def\mn@eprint@arXiv#1{\href {http://arxiv.org/abs/#1} {{\tt arXiv:#1}}}
\def\mn@eprint@dblp#1{\href {http://dblp.uni-trier.de/rec/bibtex/#1.xml}
  {dblp:#1}}
\def\mn@eprint@#1:#2:#3:#4\@nil{\def\@tempa {#1}\def\@tempb {#2}\def\@tempc
  {#3}\ifx \@tempc \@empty \let \@tempc \@tempb \let \@tempb \@tempa \fi \ifx
  \@tempb \@empty \def\@tempb {arXiv}\fi \@ifundefined
  {mn@eprint@\@tempb}{\@tempb:\@tempc}{\expandafter \expandafter \csname
  mn@eprint@\@tempb\endcsname \expandafter{\@tempc}}}

\bibitem[\protect\citeauthoryear{Abdollahi et~al.}{Abdollahi
  et~al.}{2018}]{Fermi-LAT:2018lqt}
Abdollahi S.,  et~al., 2018, \mn@doi [Science] {10.1126/science.aat8123}, 362,
  1031

\bibitem[\protect\citeauthoryear{Abe et~al.}{Abe
  et~al.}{2018}]{Hyper-Kamiokande:2018ofw}
Abe K.,  et~al., 2018, ArXiv e-print

\bibitem[\protect\citeauthoryear{Abe et~al.}{Abe
  et~al.}{2021}]{Super-Kamiokande:2021jaq}
Abe K.,  et~al., 2021, \mn@doi [Phys. Rev. D] {10.1103/PhysRevD.104.122002},
  104, 122002

\bibitem[\protect\citeauthoryear{Abe et~al.}{Abe
  et~al.}{2022}]{KamLAND:2021gvi}
Abe S.,  et~al., 2022, \mn@doi [Astrophys. J.] {10.3847/1538-4357/ac32c1}, 925,
  14

\bibitem[\protect\citeauthoryear{Adams, Kochanek, Gerke, Stanek  \& Dai}{Adams
  et~al.}{2017a}]{Adams:2016ffj}
Adams S.,  Kochanek C.,  Gerke J.,  Stanek K.,   Dai X.,  2017a, \mn@doi [Mon.
  Not. Roy. Astron. Soc.] {10.1093/mnras/stx816}, 468, 4968

\bibitem[\protect\citeauthoryear{Adams, Kochanek, Gerke  \& Stanek}{Adams
  et~al.}{2017b}]{Adams:2016hit}
Adams S.,  Kochanek C.,  Gerke J.,   Stanek K.,  2017b, \mn@doi [Mon. Not. Roy.
  Astron. Soc.] {10.1093/mnras/stx898}, 469, 1445

\bibitem[\protect\citeauthoryear{Ade et~al.,}{Ade et~al.}{2016}]{Planck2016}
Ade P. A.~R.,  et~al., 2016, \mn@doi [Astronomy \& Astrophysics]
  {10.1051/0004-6361/201525830}, 594, A13

\bibitem[\protect\citeauthoryear{An et~al.}{An et~al.}{2016}]{JUNO:2015zny}
An F.,  et~al., 2016, \mn@doi [J. Phys. G] {10.1088/0954-3899/43/3/030401}, 43,
  030401

\bibitem[\protect\citeauthoryear{Ando}{Ando}{2003}]{Ando:2003ie}
Ando S.,  2003, \mn@doi [Phys. Lett. B] {10.1016/j.physletb.2003.07.009}, 570,
  11

\bibitem[\protect\citeauthoryear{Ashida \& Nakazato}{Ashida \&
  Nakazato}{2022}]{Ashida:2022nnv}
Ashida Y.,  Nakazato K.,  2022, ArXiv e-print

\bibitem[\protect\citeauthoryear{Baker, Goldberg, Perez  \& Sarcevic}{Baker
  et~al.}{2007}]{Baker:2006gm}
Baker J.,  Goldberg H.,  Perez G.,   Sarcevic I.,  2007, \mn@doi [Phys. Rev. D]
  {10.1103/PhysRevD.76.063004}, 76, 063004

\bibitem[\protect\citeauthoryear{{Baldry} \& {Glazebrook}}{{Baldry} \&
  {Glazebrook}}{2003}]{2003ApJ...593..258B}
{Baldry} I.~K.,  {Glazebrook} K.,  2003, \mn@doi [Astrophysical Journal]
  {10.1086/376502}, \href
  {https://ui.adsabs.harvard.edu/abs/2003ApJ...593..258B} {593, 258}

\bibitem[\protect\citeauthoryear{Bays et~al.}{Bays et~al.}{2012}]{Bays:2011si}
Bays K.,  et~al., 2012, \mn@doi [Phys. Rev. D] {10.1103/PhysRevD.85.052007},
  85, 052007

\bibitem[\protect\citeauthoryear{Beacom}{Beacom}{2010}]{Beacom2010}
Beacom J.~F.,  2010, \mn@doi [Annual Review of Nuclear and Particle Science]
  {10.1146/annurev.nucl.010909.083331}, 60, 439–462

\bibitem[\protect\citeauthoryear{Beacom \& Vagins}{Beacom \&
  Vagins}{2004}]{Beacom:2003nk}
Beacom J.~F.,  Vagins M.~R.,  2004, \mn@doi [Phys. Rev. Lett.]
  {10.1103/PhysRevLett.93.171101}, 93, 171101

\bibitem[\protect\citeauthoryear{Beacom et~al.}{Beacom
  et~al.}{2017}]{Jinping:2016iiq}
Beacom J.~F.,  et~al., 2017, \mn@doi [Chin. Phys. C]
  {10.1088/1674-1137/41/2/023002}, 41, 023002

\bibitem[\protect\citeauthoryear{{Bethe} \& {Wilson}}{{Bethe} \&
  {Wilson}}{1985}]{1985ApJ...295...14B}
{Bethe} H.~A.,  {Wilson} J.~R.,  1985, \mn@doi [Astrophysical Journal]
  {10.1086/163343}, \href
  {https://ui.adsabs.harvard.edu/abs/1985ApJ...295...14B} {295, 14}

\bibitem[\protect\citeauthoryear{Bethe, Brown, Applegate  \& Lattimer}{Bethe
  et~al.}{1979a}]{Bethe:1979zd}
Bethe H.~A.,  Brown G.~E.,  Applegate J.,   Lattimer J.~M.,  1979a, \mn@doi
  [Nucl. Phys. A] {10.1016/0375-9474(79)90596-7}, 324, 487

\bibitem[\protect\citeauthoryear{{Bethe}, {Brown}, {Applegate}  \&
  {Lattimer}}{{Bethe} et~al.}{1979b}]{1979NuPhA.324..487B}
{Bethe} H.~A.,  {Brown} G.~E.,  {Applegate} J.,   {Lattimer} J.~M.,  1979b,
  \mn@doi [Nuclear Physics A] {10.1016/0375-9474(79)90596-7}, \href
  {https://ui.adsabs.harvard.edu/abs/1979NuPhA.324..487B} {324, 487}

\bibitem[\protect\citeauthoryear{{Bisnovatyi-Kogan} \&
  {Seidov}}{{Bisnovatyi-Kogan} \& {Seidov}}{1984}]{1984NYASA.422..319B}
{Bisnovatyi-Kogan} G.~S.,  {Seidov} Z.~F.,  1984, \mn@doi [Annals of the New
  York Academy of Sciences] {10.1111/j.1749-6632.1984.tb23362.x}, \href
  {https://ui.adsabs.harvard.edu/abs/1984NYASA.422..319B} {422, 319}

\bibitem[\protect\citeauthoryear{Biteau \& Williams}{Biteau \&
  Williams}{2015}]{BW2015}
Biteau J.,  Williams D.~A.,  2015, \mn@doi [The Astrophysical Journal]
  {10.1088/0004-637x/812/1/60}, 812, 60

\bibitem[\protect\citeauthoryear{Branch \& Wheeler}{Branch \&
  Wheeler}{2017}]{BranchWheeler}
Branch D.,  Wheeler J.~C.,  2017, Supernova Explosions.
Springer

\bibitem[\protect\citeauthoryear{Burrows \& Vartanyan}{Burrows \&
  Vartanyan}{2021}]{Burrows:2020qrp}
Burrows A.,  Vartanyan D.,  2021, \mn@doi [Nature]
  {10.1038/s41586-020-03059-w}, 589, 29

\bibitem[\protect\citeauthoryear{Byrne \& Fraser}{Byrne \&
  Fraser}{2022}]{Byrne:2022oik}
Byrne R.,  Fraser M.,  2022, ArXiv e-print

\bibitem[\protect\citeauthoryear{{Calzetti}}{{Calzetti}}{2013}]{2013seg..book..419C}
{Calzetti} D.,  2013, {Star Formation Rate Indicators}.
p.~419

\bibitem[\protect\citeauthoryear{{Cappellari} et~al.,}{{Cappellari}
  et~al.}{2012}]{2012Natur.484..485C}
{Cappellari} M.,  et~al., 2012, \mn@doi [{Nature}] {10.1038/nature10972}, \href
  {https://ui.adsabs.harvard.edu/abs/2012Natur.484..485C} {484, 485}

\bibitem[\protect\citeauthoryear{Cappellaro et~al.,}{Cappellaro
  et~al.}{2015}]{Cappellaro2015}
Cappellaro E.,  et~al., 2015, \mn@doi [Astronomy \& Astrophysics]
  {10.1051/0004-6361/201526712}, 584, A62

\bibitem[\protect\citeauthoryear{{Cardelli}, {Clayton}  \& {Mathis}}{{Cardelli}
  et~al.}{1989}]{1989ApJ...345..245C}
{Cardelli} J.~A.,  {Clayton} G.~C.,   {Mathis} J.~S.,  1989, \mn@doi
  [Astrophysical Journal] {10.1086/167900}, \href
  {https://ui.adsabs.harvard.edu/abs/1989ApJ...345..245C} {345, 245}

\bibitem[\protect\citeauthoryear{{Chabrier}}{{Chabrier}}{2003}]{2003PASP..115..763C}
{Chabrier} G.,  2003, \mn@doi [Publ. Astron. Soc. Pac.] {10.1086/376392}, \href
  {https://ui.adsabs.harvard.edu/abs/2003PASP..115..763C} {115, 763}

\bibitem[\protect\citeauthoryear{Chabrier, Hennebelle  \& Charlot}{Chabrier
  et~al.}{2014}]{chabrier2014}
Chabrier G.,  Hennebelle P.,   Charlot S.,  2014, \mn@doi [The Astrophysical
  Journal] {10.1088/0004-637x/796/2/75}, 796, 75

\bibitem[\protect\citeauthoryear{Chakraborty, Hansen, Izaguirre  \&
  Raffelt}{Chakraborty et~al.}{2016}]{Chakraborty:2016yeg}
Chakraborty S.,  Hansen R.,  Izaguirre I.,   Raffelt G.,  2016, \mn@doi [Nucl.
  Phys. B] {10.1016/j.nuclphysb.2016.02.012}, 908, 366

\bibitem[\protect\citeauthoryear{Chandrasekhar}{Chandrasekhar}{1931}]{Chandrasekhar:1931ih}
Chandrasekhar S.,  1931, \mn@doi [Astrophys. J.] {10.1086/143324}, 74, 81

\bibitem[\protect\citeauthoryear{{Chomiuk} \& {Povich}}{{Chomiuk} \&
  {Povich}}{2011}]{2011AJ....142..197C}
{Chomiuk} L.,  {Povich} M.~S.,  2011, \mn@doi [The Astronomical Journal]
  {10.1088/0004-6256/142/6/197}, \href
  {https://ui.adsabs.harvard.edu/abs/2011AJ....142..197C} {142, 197}

\bibitem[\protect\citeauthoryear{{Colgate} \& {White}}{{Colgate} \&
  {White}}{1966}]{1966ApJ...143..626C}
{Colgate} S.~A.,  {White} R.~H.,  1966, \mn@doi [Astrophysical Journal]
  {10.1086/148549}, \href
  {https://ui.adsabs.harvard.edu/abs/1966ApJ...143..626C} {143, 626}

\bibitem[\protect\citeauthoryear{{Couch}}{{Couch}}{2017}]{2017RSPTA.37560271C}
{Couch} S.~M.,  2017, \mn@doi [Philosophical Transactions of the Royal Society
  of London Series A] {10.1098/rsta.2016.0271}, \href
  {https://ui.adsabs.harvard.edu/abs/2017RSPTA.37560271C} {375, 20160271}

\bibitem[\protect\citeauthoryear{Creque-Sarbinowski, Hyde  \&
  Kamionkowski}{Creque-Sarbinowski et~al.}{2021}]{Creque-Sarbinowski:2020qhz}
Creque-Sarbinowski C.,  Hyde J.,   Kamionkowski M.,  2021, \mn@doi [Phys. Rev.
  D] {10.1103/PhysRevD.103.023527}, 103, 023527

\bibitem[\protect\citeauthoryear{Dahlen, Strolger, Riess, Mattila, Kankare  \&
  Mobasher}{Dahlen et~al.}{2012}]{Dahlen2012}
Dahlen T.,  Strolger L.-G.,  Riess A.~G.,  Mattila S.,  Kankare E.,   Mobasher
  B.,  2012, \mn@doi [The Astrophysical Journal] {10.1088/0004-637x/757/1/70},
  757, 70

\bibitem[\protect\citeauthoryear{Das \& Sen}{Das \& Sen}{2021}]{Das2021}
Das A.,  Sen M.,  2021, \mn@doi [Phys. Rev. D] {10.1103/PhysRevD.104.075029},
  104, 075029

\bibitem[\protect\citeauthoryear{Davies \& Beasor}{Davies \&
  Beasor}{2020}]{Davies:2020iom}
Davies B.,  Beasor E.,  2020, \mn@doi [Mon. Not. Roy. Astron. Soc.]
  {10.1093/mnrasl/slaa102}, 496, L142

\bibitem[\protect\citeauthoryear{{Driver} et~al.,}{{Driver}
  et~al.}{2011}]{2011MNRAS.413..971D}
{Driver} S.~P.,  et~al., 2011, \mn@doi [Mon. Not. Roy. Astron. Soc.]
  {10.1111/j.1365-2966.2010.18188.x}, \href
  {https://ui.adsabs.harvard.edu/abs/2011MNRAS.413..971D} {413, 971}

\bibitem[\protect\citeauthoryear{{Driver} et~al.,}{{Driver}
  et~al.}{2018}]{2018MNRAS.475.2891D}
{Driver} S.~P.,  et~al., 2018, \mn@doi [Mon. Not. Roy. Astron. Soc.]
  {10.1093/mnras/stx2728}, \href
  {https://ui.adsabs.harvard.edu/abs/2018MNRAS.475.2891D} {475, 2891}

\bibitem[\protect\citeauthoryear{Duan, Fuller  \& Qian}{Duan
  et~al.}{2010}]{Duan:2010bg}
Duan H.,  Fuller G.~M.,   Qian Y.-Z.,  2010, \mn@doi [Ann. Rev. Nucl. Part.
  Sci.] {10.1146/annurev.nucl.012809.104524}, 60, 569

\bibitem[\protect\citeauthoryear{El~Hedri, Ashida  \& Giampaolo}{El~Hedri
  et~al.}{2021}]{elhedri:hal-03373391}
El~Hedri S.,  Ashida Y.,   Giampaolo A.,  2021, in {37th International Cosmic
  Ray Conference}. Berlin, Germany, p.~1139, \mn@doi{10.22323/1.395.1139}, \url
  {https://hal.archives-ouvertes.fr/hal-03373391}

\bibitem[\protect\citeauthoryear{Ertl, Janka, Woosley, Sukhbold  \&
  Ugliano}{Ertl et~al.}{2016}]{Ertl:2015rga}
Ertl T.,  Janka H.~T.,  Woosley S.,  Sukhbold T.,   Ugliano M.,  2016, \mn@doi
  [Astrophys. J.] {10.3847/0004-637X/818/2/124}, 818, 124

\bibitem[\protect\citeauthoryear{{Fardal}, {Katz}, {Weinberg}  \&
  {Dav{\'e}}}{{Fardal} et~al.}{2007}]{2007MNRAS.379..985F}
{Fardal} M.~A.,  {Katz} N.,  {Weinberg} D.~H.,   {Dav{\'e}} R.,  2007, \mn@doi
  [Mon. Not. Roy. Astron. Soc.] {10.1111/j.1365-2966.2007.11522.x}, \href
  {https://ui.adsabs.harvard.edu/abs/2007MNRAS.379..985F} {379, 985}

\bibitem[\protect\citeauthoryear{Farzan \& Palomares-Ruiz}{Farzan \&
  Palomares-Ruiz}{2014}]{Farzan:2014gza}
Farzan Y.,  Palomares-Ruiz S.,  2014, \mn@doi [JCAP]
  {10.1088/1475-7516/2014/06/014}, 06, 014

\bibitem[\protect\citeauthoryear{Ferreras, Barbera, Rosa, Vazdekis, Carvalho,
  Falcón-Barroso  \& Ricciardelli}{Ferreras et~al.}{2012}]{Ferreras2012}
Ferreras I.,  Barbera F.~L.,  Rosa I. G. d.~l.,  Vazdekis A.,  Carvalho R.
  R.~d.,  Falcón-Barroso J.,   Ricciardelli E.,  2012, \mn@doi [Monthly
  Notices of the Royal Astronomical Society: Letters] {10.1093/mnrasl/sls014},
  429, L15–L19

\bibitem[\protect\citeauthoryear{Ferré-Mateu, Vazdekis  \& de~la
  Rosa}{Ferré-Mateu et~al.}{2013}]{FM2013}
Ferré-Mateu A.,  Vazdekis A.,   de~la Rosa I.~G.,  2013, \mn@doi [Monthly
  Notices of the Royal Astronomical Society] {10.1093/mnras/stt193}, 431,
  440–454

\bibitem[\protect\citeauthoryear{{Fioc} \& {Rocca-Volmerange}}{{Fioc} \&
  {Rocca-Volmerange}}{2019}]{2019A&A...623A.143F}
{Fioc} M.,  {Rocca-Volmerange} B.,  2019, \mn@doi [Astronomy \& Astrophysics]
  {10.1051/0004-6361/201833556}, \href
  {https://ui.adsabs.harvard.edu/abs/2019A&A...623A.143F} {623, A143}

\bibitem[\protect\citeauthoryear{Fishbach, Holz  \& Farr}{Fishbach
  et~al.}{2018}]{Fishbach:2018edt}
Fishbach M.,  Holz D.~E.,   Farr W.~M.,  2018, \mn@doi [Astrophys. J. Lett.]
  {10.3847/2041-8213/aad800}, 863, L41

\bibitem[\protect\citeauthoryear{Fogli, Lisi, Mirizzi  \& Montanino}{Fogli
  et~al.}{2004}]{Fogli:2004gy}
Fogli G.~L.,  Lisi E.,  Mirizzi A.,   Montanino D.,  2004, \mn@doi [Phys. Rev.
  D] {10.1103/PhysRevD.70.013001}, 70, 013001

\bibitem[\protect\citeauthoryear{{Fontanot}, {De Lucia}, {Hirschmann},
  {Bruzual}, {Charlot}  \& {Zibetti}}{{Fontanot}
  et~al.}{2017}]{2017MNRAS.464.3812F}
{Fontanot} F.,  {De Lucia} G.,  {Hirschmann} M.,  {Bruzual} G.,  {Charlot} S.,
   {Zibetti} S.,  2017, \mn@doi [Mon. Not. Roy. Astron. Soc.]
  {10.1093/mnras/stw2612}, \href
  {https://ui.adsabs.harvard.edu/abs/2017MNRAS.464.3812F} {464, 3812}

\bibitem[\protect\citeauthoryear{Fontanot, Lucia, Xie, Hirschmann, Bruzual  \&
  Charlot}{Fontanot et~al.}{2018}]{Fontanot_2018}
Fontanot F.,  Lucia G.~D.,  Xie L.,  Hirschmann M.,  Bruzual G.,   Charlot S.,
  2018, \mn@doi [Monthly Notices of the Royal Astronomical Society]
  {10.1093/mnras/stx3323}

\bibitem[\protect\citeauthoryear{Freese, Rindler-Daller, Spolyar  \&
  Valluri}{Freese et~al.}{2016}]{freese2016}
Freese K.,  Rindler-Daller T.,  Spolyar D.,   Valluri M.,  2016, \mn@doi
  [Reports on Progress in Physics] {10.1088/0034-4885/79/6/066902}, 79

\bibitem[\protect\citeauthoryear{Fuller}{Fuller}{1982}]{Fuller:1981mu}
Fuller G.~M.,  1982, \mn@doi [Astrophys. J.] {10.1086/159598}, 252, 741

\bibitem[\protect\citeauthoryear{{Garching Core-Collapse Supernova
  Archive}}{{Garching Core-Collapse Supernova Archive}}{2022}]{Garc:SN}
{Garching Core-Collapse Supernova Archive} 2022, {},
  \url{https://wwwmpa.mpa-garching.mpg.de/ccsnarchive/}

\bibitem[\protect\citeauthoryear{{Geha}}{{Geha}}{2013}]{2013hst..prop13449G}
{Geha} M.,  2013, {A Non-Universal Initial Mass Function in the Ultra-Faint
  Galaxy Coma Berenices}, HST Proposal ID 13449. Cycle 21

\bibitem[\protect\citeauthoryear{Gerke, Kochanek  \& Stanek}{Gerke
  et~al.}{2015}]{Gerke:2014ooa}
Gerke J.,  Kochanek C.,   Stanek K.,  2015, \mn@doi [Mon. Not. Roy. Astron.
  Soc.] {10.1093/mnras/stv776}, 450, 3289

\bibitem[\protect\citeauthoryear{Goldberg, Perez  \& Sarcevic}{Goldberg
  et~al.}{2006}]{Goldberg:2005yw}
Goldberg H.,  Perez G.,   Sarcevic I.,  2006, \mn@doi [JHEP]
  {10.1088/1126-6708/2006/11/023}, 11, 023

\bibitem[\protect\citeauthoryear{Gruppioni et~al.}{Gruppioni
  et~al.}{2013}]{Gruppioni:2013jna}
Gruppioni C.,  et~al., 2013, \mn@doi [Mon. Not. Roy. Astron. Soc.]
  {10.1093/mnras/stt308}, 432, 23

\bibitem[\protect\citeauthoryear{{Gunawardhana} et~al.,}{{Gunawardhana}
  et~al.}{2011}]{2011MNRAS.415.1647G}
{Gunawardhana} M.~L.~P.,  et~al., 2011, \mn@doi [Monthly Notices of the Royal
  Astronomical Society] {10.1111/j.1365-2966.2011.18800.x}, \href
  {https://ui.adsabs.harvard.edu/abs/2011MNRAS.415.1647G} {415, 1647}

\bibitem[\protect\citeauthoryear{Harayama, Eisenhauer  \& Martins}{Harayama
  et~al.}{2008}]{Harayama2008}
Harayama Y.,  Eisenhauer F.,   Martins F.,  2008, \mn@doi [The Astrophysical
  Journal] {10.1086/524650}, 675, 1319–1342

\bibitem[\protect\citeauthoryear{Heger, Fryer, Woosley, Langer  \&
  Hartmann}{Heger et~al.}{2003}]{Heger:2002by}
Heger A.,  Fryer C.~L.,  Woosley S.~E.,  Langer N.,   Hartmann D.~H.,  2003,
  \mn@doi [Astrophys. J.] {10.1086/375341}, 591, 288

\bibitem[\protect\citeauthoryear{Hopkins}{Hopkins}{2018}]{Hopkins2018}
Hopkins A.~M.,  2018, \mn@doi [Publications of the Astronomical Society of
  Australia] {10.1017/pasa.2018.29}, 35

\bibitem[\protect\citeauthoryear{{Hopkins} \& {Beacom}}{{Hopkins} \&
  {Beacom}}{2006}]{2006ApJ...651..142H}
{Hopkins} A.~M.,  {Beacom} J.~F.,  2006, \mn@doi [Astrophysical Journal]
  {10.1086/506610}, \href
  {https://ui.adsabs.harvard.edu/abs/2006ApJ...651..142H} {651, 142}

\bibitem[\protect\citeauthoryear{Horiuchi, Beacom  \& Dwek}{Horiuchi
  et~al.}{2009}]{Horiuchi:2008jz}
Horiuchi S.,  Beacom J.~F.,   Dwek E.,  2009, \mn@doi [Phys. Rev. D]
  {10.1103/PhysRevD.79.083013}, 79, 083013

\bibitem[\protect\citeauthoryear{Horiuchi, Beacom, Kochanek, Prieto, Stanek  \&
  Thompson}{Horiuchi et~al.}{2011}]{Horiuchi:2011zz}
Horiuchi S.,  Beacom J.~F.,  Kochanek C.~S.,  Prieto J.~L.,  Stanek K.~Z.,
  Thompson T.~A.,  2011, \mn@doi [Astrophys. J.] {10.1088/0004-637X/738/2/154},
  738, 154

\bibitem[\protect\citeauthoryear{Horiuchi, Beacom, Bothwell  \&
  Thompson}{Horiuchi et~al.}{2013}]{Horiuchi:2013bc}
Horiuchi S.,  Beacom J.~F.,  Bothwell M.~S.,   Thompson T.~A.,  2013, \mn@doi
  [Astrophys. J.] {10.1088/0004-637X/769/2/113}, 769, 113

\bibitem[\protect\citeauthoryear{Horiuchi, Nakamura, Takiwaki, Kotake  \&
  Tanaka}{Horiuchi et~al.}{2014}]{Horiuchi:2014ska}
Horiuchi S.,  Nakamura K.,  Takiwaki T.,  Kotake K.,   Tanaka M.,  2014,
  \mn@doi [Mon. Not. Roy. Astron. Soc.] {10.1093/mnrasl/slu146}, 445, L99

\bibitem[\protect\citeauthoryear{Horiuchi, Sumiyoshi, Nakamura, Fischer, Summa,
  Takiwaki, Janka  \& Kotake}{Horiuchi et~al.}{2018}]{Horiuchi:2017qja}
Horiuchi S.,  Sumiyoshi K.,  Nakamura K.,  Fischer T.,  Summa A.,  Takiwaki T.,
   Janka H.-T.,   Kotake K.,  2018, \mn@doi [Mon. Not. Roy. Astron. Soc.]
  {10.1093/mnras/stx3271}, 475, 1363

\bibitem[\protect\citeauthoryear{Horiuchi, Kinugawa, Takiwaki, Takahashi  \&
  Kotake}{Horiuchi et~al.}{2021}]{Horiuchi:2020jnc}
Horiuchi S.,  Kinugawa T.,  Takiwaki T.,  Takahashi K.,   Kotake K.,  2021,
  \mn@doi [Phys. Rev. D] {10.1103/PhysRevD.103.043003}, 103, 043003

\bibitem[\protect\citeauthoryear{{Ivezi{\'c}} et~al.,}{{Ivezi{\'c}}
  et~al.}{2019}]{2019ApJ...873..111I}
{Ivezi{\'c}} {\v Z}.,  et~al., 2019, \mn@doi [Astrophysical Journal]
  {10.3847/1538-4357/ab042c}, \href
  {http://adsabs.harvard.edu/abs/2019ApJ...873..111I} {873, 111}

\bibitem[\protect\citeauthoryear{Janka}{Janka}{2017}]{Janka:2017vcp}
Janka H.-T.,  2017, Neutrino-Driven Explosions.
Springer International Publishing, Cham, pp 1095--1150,
  \mn@doi{10.1007/978-3-319-21846-5_109}, \url
  {https://doi.org/10.1007/978-3-319-21846-5_109}

\bibitem[\protect\citeauthoryear{Jeong, Palomares-Ruiz, Reno  \&
  Sarcevic}{Jeong et~al.}{2018}]{Jeong:2018yts}
Jeong Y.~S.,  Palomares-Ruiz S.,  Reno M.~H.,   Sarcevic I.,  2018, \mn@doi
  [JCAP] {10.1088/1475-7516/2018/06/019}, 06, 019

\bibitem[\protect\citeauthoryear{Keehn \& Lunardini}{Keehn \&
  Lunardini}{2012}]{Keehn:2010pn}
Keehn J.~G.,  Lunardini C.,  2012, \mn@doi [Phys. Rev. D]
  {10.1103/PhysRevD.85.043011}, 85, 043011

\bibitem[\protect\citeauthoryear{Keil}{Keil}{2003}]{Keil:2003sw}
Keil M.~T.,  2003, Other thesis (\mn@eprint {arXiv} {astro-ph/0308228})

\bibitem[\protect\citeauthoryear{Keil, Raffelt  \& Janka}{Keil
  et~al.}{2003}]{Keil:2002in}
Keil M.~T.,  Raffelt G.~G.,   Janka H.-T.,  2003, \mn@doi [Astrophys. J.]
  {10.1086/375130}, 590, 971

\bibitem[\protect\citeauthoryear{Kennedy, Wyatt, Kalas, Duchene, Sibthorpe,
  Lestrade, Matthews  \& Greaves}{Kennedy et~al.}{2013}]{Kennedy_2013}
Kennedy G.~M.,  Wyatt M.~C.,  Kalas P.,  Duchene G.,  Sibthorpe B.,  Lestrade
  J.-F.,  Matthews B.~C.,   Greaves J.,  2013, \mn@doi [Monthly Notices of the
  Royal Astronomical Society: Letters] {10.1093/mnrasl/slt168}, 438, L96

\bibitem[\protect\citeauthoryear{{Kennicutt}}{{Kennicutt}}{1998}]{1998ARA&A..36..189K}
{Kennicutt} Robert~C. J.,  1998, \mn@doi [Annual Review of Astronomy and
  Astrophysics] {10.1146/annurev.astro.36.1.189}, \href
  {https://ui.adsabs.harvard.edu/abs/1998ARA&A..36..189K} {36, 189}

\bibitem[\protect\citeauthoryear{Kennicutt \& Evans}{Kennicutt \&
  Evans}{2012}]{KE2012}
Kennicutt R.~C.,  Evans N.~J.,  2012, \mn@doi [Annual Review of Astronomy and
  Astrophysics] {10.1146/annurev-astro-081811-125610}, 50, 531–608

\bibitem[\protect\citeauthoryear{{Kobayashi}}{{Kobayashi}}{2010}]{2010AIPC.1240..123K}
{Kobayashi} C.,  2010, in {Debattista} V.~P.,  {Popescu} C.~C.,  eds,  American
  Institute of Physics Conference Series Vol. 1240, Hunting for the Dark: the
  Hidden Side of Galaxy Formation. pp 123--126 (\mn@eprint {arXiv}
  {1002.4475}), \mn@doi{10.1063/1.3458465}

\bibitem[\protect\citeauthoryear{Kochanek, Beacom, Kistler, Prieto, Stanek,
  Thompson  \& Yuksel}{Kochanek et~al.}{2008}]{Kochanek:2008mp}
Kochanek C.,  Beacom J.,  Kistler M.,  Prieto J.,  Stanek K.,  Thompson T.,
  Yuksel H.,  2008, \mn@doi [Astrophys. J.] {10.1086/590053}, 684, 1336

\bibitem[\protect\citeauthoryear{Koekemoer et~al.,}{Koekemoer
  et~al.}{2019}]{koekemoer2019ultra}
Koekemoer A.~M.,  et~al., 2019, An Ultra Deep Field survey with WFIRST
  (\mn@eprint {arXiv} {1903.06154})

\bibitem[\protect\citeauthoryear{Krauss, Glashow  \& Schramm}{Krauss
  et~al.}{1984}]{Krauss:1983zn}
Krauss L.~M.,  Glashow S.~L.,   Schramm D.~N.,  1984, \mn@doi [Nature]
  {10.1038/310191a0}, 310, 191

\bibitem[\protect\citeauthoryear{Kresse, Ertl  \& Janka}{Kresse
  et~al.}{2021}]{Kresse:2020nto}
Kresse D.,  Ertl T.,   Janka H.-T.,  2021, \mn@doi [Astrophys. J.]
  {10.3847/1538-4357/abd54e}, 909, 169

\bibitem[\protect\citeauthoryear{{Kroupa}}{{Kroupa}}{2001}]{2001MNRAS.322..231K}
{Kroupa} P.,  2001, \mn@doi [Mon. Not. Roy. Astron. Soc.]
  {10.1046/j.1365-8711.2001.04022.x}, \href
  {https://ui.adsabs.harvard.edu/abs/2001MNRAS.322..231K} {322, 231}

\bibitem[\protect\citeauthoryear{La~Barbera et~al.,}{La~Barbera
  et~al.}{2019}]{LaBarbera2019}
La~Barbera F.,  et~al., 2019, \mn@doi [Monthly Notices of the Royal
  Astronomical Society] {10.1093/mnras/stz2192}, 489, 4090–4110

\bibitem[\protect\citeauthoryear{{Larson}}{{Larson}}{1998}]{1998MNRAS.301..569L}
{Larson} R.~B.,  1998, \mn@doi [Mon. Not. Roy. Astron. Soc.]
  {10.1046/j.1365-8711.1998.02045.x}, \href
  {https://ui.adsabs.harvard.edu/abs/1998MNRAS.301..569L} {301, 569}

\bibitem[\protect\citeauthoryear{Lattimer \& Swesty}{Lattimer \&
  Swesty}{1991}]{Lattimer:1991nc}
Lattimer J.~M.,  Swesty F.~D.,  1991, \mn@doi [Nucl. Phys.]
  {10.1016/0375-9474(91)90452-C}, A535, 331

\bibitem[\protect\citeauthoryear{Laureijs et~al.,}{Laureijs
  et~al.}{2011}]{laureijs2011euclid}
Laureijs R.,  et~al., 2011, Euclid Definition Study Report (\mn@eprint {arXiv}
  {1110.3193})

\bibitem[\protect\citeauthoryear{Leitherer et~al.,}{Leitherer
  et~al.}{1999}]{Leitherer:1999rq}
Leitherer C.,  et~al., 1999, \mn@doi [Astrophys. J. Suppl.] {10.1086/313233},
  123, 3

\bibitem[\protect\citeauthoryear{{Li} \& {Draine}}{{Li} \&
  {Draine}}{2001}]{2001ApJ...554..778L}
{Li} A.,  {Draine} B.~T.,  2001, \mn@doi [The Astrophysical Journal]
  {10.1086/323147}, \href
  {https://ui.adsabs.harvard.edu/abs/2001ApJ...554..778L} {554, 778}

\bibitem[\protect\citeauthoryear{Li, Vagins  \& Wurm}{Li
  et~al.}{2022}]{Li:2022myd}
Li Y.-F.,  Vagins M.,   Wurm M.,  2022, \mn@doi [Universe]
  {10.3390/universe8030181}, 8, 181

\bibitem[\protect\citeauthoryear{Libanov \& Sharofeev}{Libanov \&
  Sharofeev}{2022}]{Libanov:2022yta}
Libanov A.,  Sharofeev A.,  2022, ArXiv e-print

\bibitem[\protect\citeauthoryear{Lien, Fields  \& Beacom}{Lien
  et~al.}{2010}]{Lien:2010yb}
Lien A.,  Fields B.~D.,   Beacom J.~F.,  2010, \mn@doi [Phys. Rev.]
  {10.1103/PhysRevD.81.083001}, D81, 083001

\bibitem[\protect\citeauthoryear{Lunardini}{Lunardini}{2009}]{Lunardini:2009ya}
Lunardini C.,  2009, \mn@doi [Phys. Rev. Lett.]
  {10.1103/PhysRevLett.102.231101}, 102, 231101

\bibitem[\protect\citeauthoryear{Lunardini}{Lunardini}{2016}]{Lunardini:2010ab}
Lunardini C.,  2016, \mn@doi [Astropart. Phys.]
  {10.1016/j.astropartphys.2016.02.005}, 79, 49

\bibitem[\protect\citeauthoryear{Lunardini \& Tamborra}{Lunardini \&
  Tamborra}{2012}]{Lunardini:2012ne}
Lunardini C.,  Tamborra I.,  2012, \mn@doi [JCAP]
  {10.1088/1475-7516/2012/07/012}, 07, 012

\bibitem[\protect\citeauthoryear{Madau \& Dickinson}{Madau \&
  Dickinson}{2014}]{Madau:2014bja}
Madau P.,  Dickinson M.,  2014, \mn@doi [Ann. Rev. Astron. Astrophys.]
  {10.1146/annurev-astro-081811-125615}, 52, 415

\bibitem[\protect\citeauthoryear{{Magnelli}, {Elbaz}, {Chary}, {Dickinson}, {Le
  Borgne}, {Frayer}  \& {Willmer}}{{Magnelli}
  et~al.}{2011}]{2011A&A...528A..35M}
{Magnelli} B.,  {Elbaz} D.,  {Chary} R.~R.,  {Dickinson} M.,  {Le Borgne} D.,
  {Frayer} D.~T.,   {Willmer} C.~N.~A.,  2011, \mn@doi [Astronomy \&
  Astrophysics] {10.1051/0004-6361/200913941}, \href
  {https://ui.adsabs.harvard.edu/abs/2011A&A...528A..35M} {528, A35}

\bibitem[\protect\citeauthoryear{{Magnelli} et~al.,}{{Magnelli}
  et~al.}{2013}]{2013A&A...553A.132M}
{Magnelli} B.,  et~al., 2013, \mn@doi [Astronomy \& Astrophysics]
  {10.1051/0004-6361/201321371}, \href
  {https://ui.adsabs.harvard.edu/abs/2013A&A...553A.132M} {553, A132}

\bibitem[\protect\citeauthoryear{Malek et~al.,}{Malek
  et~al.}{2003}]{SuperKamiokande2003}
Malek M.,  et~al., 2003, \mn@doi [Physical Review Letters]
  {10.1103/physrevlett.90.061101}, 90

\bibitem[\protect\citeauthoryear{Maoz, Mannucci  \& Nelemans}{Maoz
  et~al.}{2014}]{Maoz2014}
Maoz D.,  Mannucci F.,   Nelemans G.,  2014, \mn@doi [Annual Review of
  Astronomy and Astrophysics] {10.1146/annurev-astro-082812-141031}, 52,
  107–170

\bibitem[\protect\citeauthoryear{Mathews, Hidaka, Kajino  \& Suzuki}{Mathews
  et~al.}{2014}]{Mathews:2014qba}
Mathews G.~J.,  Hidaka J.,  Kajino T.,   Suzuki J.,  2014, \mn@doi [Astrophys.
  J.] {10.1088/0004-637X/790/2/115}, 790, 115

\bibitem[\protect\citeauthoryear{Mattila et~al.,}{Mattila
  et~al.}{2012}]{Mattila2012}
Mattila S.,  et~al., 2012, \mn@doi [The Astrophysical Journal]
  {10.1088/0004-637x/756/2/111}, 756, 111

\bibitem[\protect\citeauthoryear{Mazzali, R{\:o}pke, Benetti  \&
  Hillebrandt}{Mazzali et~al.}{2007}]{Mazzali2007}
Mazzali P.~A.,  R{\:o}pke F.~K.,  Benetti S.,   Hillebrandt W.,  2007, \mn@doi
  [Science] {10.1126/science.1136259}, 315, 825–828

\bibitem[\protect\citeauthoryear{Mirizzi, Tamborra, Janka, Saviano, Scholberg,
  Bollig, Hudepohl  \& Chakraborty}{Mirizzi et~al.}{2016}]{Mirizzi:2015eza}
Mirizzi A.,  Tamborra I.,  Janka H.-T.,  Saviano N.,  Scholberg K.,  Bollig R.,
   Hudepohl L.,   Chakraborty S.,  2016, \mn@doi [Riv. Nuovo Cim.]
  {10.1393/ncr/i2016-10120-8}, 39, 1

\bibitem[\protect\citeauthoryear{Møller, Suliga, Tamborra  \& Denton}{Møller
  et~al.}{2018}]{Moller:2018kpn}
Møller K.,  Suliga A.~M.,  Tamborra I.,   Denton P.~B.,  2018, \mn@doi [JCAP]
  {10.1088/1475-7516/2018/05/066}, 1805, 066

\bibitem[\protect\citeauthoryear{Nakazato}{Nakazato}{2013}]{Nakazato:2013maa}
Nakazato K.,  2013, \mn@doi [Phys. Rev. D] {10.1103/PhysRevD.88.083012}, 88,
  083012

\bibitem[\protect\citeauthoryear{Nakazato, Mochida, Niino  \& Suzuki}{Nakazato
  et~al.}{2015}]{Nakazato:2015rya}
Nakazato K.,  Mochida E.,  Niino Y.,   Suzuki H.,  2015, \mn@doi [Astrophys.
  J.] {10.1088/0004-637X/804/1/75}, 804, 75

\bibitem[\protect\citeauthoryear{Neustadt, Kochanek, Stanek, Basinger,
  Jayasinghe, Garling, Adams  \& Gerke}{Neustadt
  et~al.}{2021}]{Neustadt:2021jjt}
Neustadt J. M.~M.,  Kochanek C.~S.,  Stanek K.~Z.,  Basinger C.~M.,  Jayasinghe
  T.,  Garling C.~T.,  Adams S.~M.,   Gerke J.,  2021, \mn@doi [Mon. Not. Roy.
  Astron. Soc.] {10.1093/mnras/stab2605}, 508, 516

\bibitem[\protect\citeauthoryear{Offner, Clark, Hennebelle, Bastian, Bate,
  Hopkins, Moreaux  \& Whitworth}{Offner et~al.}{2014}]{Offner2014}
Offner S. S.~R.,  Clark P.~C.,  Hennebelle P.,  Bastian N.,  Bate M.~R.,
  Hopkins P.~F.,  Moreaux E.,   Whitworth A.~P.,  2014, \mn@doi [Protostars and
  Planets VI] {10.2458/azu_uapress_9780816531240-ch003}

\bibitem[\protect\citeauthoryear{{Padoan}, {Nordlund}  \& {Jones}}{{Padoan}
  et~al.}{1997}]{1997MNRAS.288..145P}
{Padoan} P.,  {Nordlund} A.,   {Jones} B. J.~T.,  1997, \mn@doi [Mon. Not. Roy.
  Astron. Soc.] {10.1093/mnras/288.1.145}, \href
  {https://ui.adsabs.harvard.edu/abs/1997MNRAS.288..145P} {288, 145}

\bibitem[\protect\citeauthoryear{Perrett et~al.,}{Perrett
  et~al.}{2012}]{Perrett2012}
Perrett K.,  et~al., 2012, \mn@doi [The Astronomical Journal]
  {10.1088/0004-6256/144/2/59}, 144, 59

\bibitem[\protect\citeauthoryear{Petrushevska et~al.}{Petrushevska
  et~al.}{2016}]{Petrushevska:2016kie}
Petrushevska T.,  et~al., 2016, \mn@doi [Astron. Astrophys.]
  {10.1051/0004-6361/201628925}, 594, A54

\bibitem[\protect\citeauthoryear{Priya \& Lunardini}{Priya \&
  Lunardini}{2017}]{Priya:2017bmm}
Priya A.,  Lunardini C.,  2017, \mn@doi [JCAP] {10.1088/1475-7516/2017/11/031},
  11, 031

\bibitem[\protect\citeauthoryear{Razzaque, Dermer  \& Finke}{Razzaque
  et~al.}{2009}]{Razzaque2009}
Razzaque S.,  Dermer C.~D.,   Finke J.~D.,  2009, \mn@doi [The Astrophysical
  Journal] {10.1088/0004-637x/697/1/483}, 697, 483–492

\bibitem[\protect\citeauthoryear{{Reg{\H{o}}s} \& {Vink{\'o}}}{{Reg{\H{o}}s} \&
  {Vink{\'o}}}{2019}]{2019ApJ...874..158R}
{Reg{\H{o}}s} E.,  {Vink{\'o}} J.,  2019, \mn@doi [The Astrophysical Journal]
  {10.3847/1538-4357/ab0a73}, \href
  {https://ui.adsabs.harvard.edu/abs/2019ApJ...874..158R} {874, 158}

\bibitem[\protect\citeauthoryear{Rose et~al.,}{Rose
  et~al.}{2021}]{https://doi.org/10.48550/arxiv.2111.03081}
Rose B.~M.,  et~al., 2021, A Reference Survey for Supernova Cosmology with the
  Nancy Grace Roman Space Telescope, \mn@doi{10.48550/ARXIV.2111.03081}, \url
  {https://arxiv.org/abs/2111.03081}

\bibitem[\protect\citeauthoryear{{Salpeter}}{{Salpeter}}{1955}]{1955ApJ...121..161S}
{Salpeter} E.~E.,  1955, \mn@doi [Astrophysical Journal] {10.1086/145971},
  \href {https://ui.adsabs.harvard.edu/abs/1955ApJ...121..161S} {121, 161}

\bibitem[\protect\citeauthoryear{Sawatzki, Wurm  \& Kresse}{Sawatzki
  et~al.}{2021}]{Sawatzki:2020mpb}
Sawatzki J.,  Wurm M.,   Kresse D.,  2021, \mn@doi [Phys. Rev. D]
  {10.1103/PhysRevD.103.023021}, 103, 023021

\bibitem[\protect\citeauthoryear{Sharda \& Krumholz}{Sharda \&
  Krumholz}{2021}]{Sharda2021}
Sharda P.,  Krumholz M.~R.,  2021, \mn@doi [Monthly Notices of the Royal
  Astronomical Society] {10.1093/mnras/stab2921}

\bibitem[\protect\citeauthoryear{Singh \& Rentala}{Singh \&
  Rentala}{2021}]{Singh:2020tmt}
Singh R.,  Rentala V.,  2021, \mn@doi [JCAP] {10.1088/1475-7516/2021/08/019},
  08, 019

\bibitem[\protect\citeauthoryear{Smith}{Smith}{2014}]{Smith:2014txa}
Smith N.,  2014, \mn@doi [Ann. Rev. Astron. Astrophys.]
  {10.1146/annurev-astro-081913-040025}, 52, 487

\bibitem[\protect\citeauthoryear{Strolger et~al.,}{Strolger
  et~al.}{2015}]{Strolger:2015kra}
Strolger L.-G.,  et~al., 2015, \mn@doi [Astrophys. J.]
  {10.1088/0004-637X/813/2/93}, 813, 93

\bibitem[\protect\citeauthoryear{Strolger, Rodney, Pacifici, Narayan  \&
  Graur}{Strolger et~al.}{2020}]{Strolger2020}
Strolger L.-G.,  Rodney S.~A.,  Pacifici C.,  Narayan G.,   Graur O.,  2020,
  \mn@doi [The Astrophysical Journal] {10.3847/1538-4357/ab6a97}, 890, 140

\bibitem[\protect\citeauthoryear{Sukhbold, Ertl, Woosley, Brown  \&
  Janka}{Sukhbold et~al.}{2016}]{Sukhbold:2015wba}
Sukhbold T.,  Ertl T.,  Woosley S.,  Brown J.~M.,   Janka H.~T.,  2016, \mn@doi
  [Astrophys. J.] {10.3847/0004-637X/821/1/38}, 821, 38

\bibitem[\protect\citeauthoryear{Suliga, Beacom  \& Tamborra}{Suliga
  et~al.}{2022}]{Suliga:2021hek}
Suliga A.~M.,  Beacom J.~F.,   Tamborra I.,  2022, \mn@doi [Phys. Rev. D]
  {10.1103/PhysRevD.105.043008}, 105, 043008

\bibitem[\protect\citeauthoryear{Tabrizi \& Horiuchi}{Tabrizi \&
  Horiuchi}{2021}]{Tabrizi:2020vmo}
Tabrizi Z.,  Horiuchi S.,  2021, \mn@doi [JCAP]
  {10.1088/1475-7516/2021/05/011}, 05, 011

\bibitem[\protect\citeauthoryear{Tamborra \& Shalgar}{Tamborra \&
  Shalgar}{2021}]{Tamborra:2020cul}
Tamborra I.,  Shalgar S.,  2021, \mn@doi [Ann. Rev. Nucl. Part. Sci.]
  {10.1146/annurev-nucl-102920-050505}, 71, 165

\bibitem[\protect\citeauthoryear{Tamborra, Muller, Hudepohl, Janka  \&
  Raffelt}{Tamborra et~al.}{2012}]{Tamborra:2012ac}
Tamborra I.,  Muller B.,  Hudepohl L.,  Janka H.-T.,   Raffelt G.,  2012,
  \mn@doi [Phys. Rev. D] {10.1103/PhysRevD.86.125031}, 86, 125031

\bibitem[\protect\citeauthoryear{Turatto}{Turatto}{2003}]{Turatto2003}
Turatto M.,  2003, \mn@doi [Lecture Notes in Physics]
  {10.1007/3-540-45863-8_3}, p. 21–36

\bibitem[\protect\citeauthoryear{Vitale, Farr, Ng  \& Rodriguez}{Vitale
  et~al.}{2019}]{Vitale2019}
Vitale S.,  Farr W.~M.,  Ng K. K.~Y.,   Rodriguez C.~L.,  2019, \mn@doi [The
  Astrophysical Journal] {10.3847/2041-8213/ab50c0}, 886, L1

\bibitem[\protect\citeauthoryear{{Weingartner} \& {Draine}}{{Weingartner} \&
  {Draine}}{2001}]{2001ApJ...548..296W}
{Weingartner} J.~C.,  {Draine} B.~T.,  2001, \mn@doi [The Astrophysical
  Journal] {10.1086/318651}, \href
  {https://ui.adsabs.harvard.edu/abs/2001ApJ...548..296W} {548, 296}

\bibitem[\protect\citeauthoryear{{Wilkins}, {Trentham}  \& {Hopkins}}{{Wilkins}
  et~al.}{2008a}]{2008MNRAS.385..687W}
{Wilkins} S.~M.,  {Trentham} N.,   {Hopkins} A.~M.,  2008a, \mn@doi [\mnras]
  {10.1111/j.1365-2966.2008.12885.x}, \href
  {https://ui.adsabs.harvard.edu/abs/2008MNRAS.385..687W} {385, 687}

\bibitem[\protect\citeauthoryear{{Wilkins}, {Hopkins}, {Trentham}  \&
  {Tojeiro}}{{Wilkins} et~al.}{2008b}]{2008MNRAS.391..363W}
{Wilkins} S.~M.,  {Hopkins} A.~M.,  {Trentham} N.,   {Tojeiro} R.,  2008b,
  \mn@doi [\mnras] {10.1111/j.1365-2966.2008.13890.x}, \href
  {https://ui.adsabs.harvard.edu/abs/2008MNRAS.391..363W} {391, 363}

\bibitem[\protect\citeauthoryear{Wilkins, Lovell  \& Stanway}{Wilkins
  et~al.}{2019}]{Wilkins2019}
Wilkins S.~M.,  Lovell C.~C.,   Stanway E.~R.,  2019, \mn@doi [Monthly Notices
  of the Royal Astronomical Society] {10.1093/mnras/stz2894}, 490, 5359–5365

\bibitem[\protect\citeauthoryear{Wilson, Mayle, Woosley  \& Weaver}{Wilson
  et~al.}{1986}]{Wilson:1986ha}
Wilson J.~R.,  Mayle R.,  Woosley S.~E.,   Weaver T.,  1986, \mn@doi [Annals N.
  Y. Acad. Sci.] {10.1111/j.1749-6632.1986.tb47980.x}, 470, 267

\bibitem[\protect\citeauthoryear{Zhang et~al.}{Zhang
  et~al.}{2015}]{Zhang:2013tua}
Zhang H.,  et~al., 2015, \mn@doi [Astropart. Phys.]
  {10.1016/j.astropartphys.2014.05.004}, 60, 41

\bibitem[\protect\citeauthoryear{{Zubko}, {Dwek}  \& {Arendt}}{{Zubko}
  et~al.}{2004}]{2004ApJS..152..211Z}
{Zubko} V.,  {Dwek} E.,   {Arendt} R.~G.,  2004, \mn@doi [The Astrophysical
  Journal Supplement Series] {10.1086/382351}, \href
  {https://ui.adsabs.harvard.edu/abs/2004ApJS..152..211Z} {152, 211}

\bibitem[\protect\citeauthoryear{de Gouv\^ea, Martinez-Soler, Perez-Gonzalez
  \& Sen}{de~Gouv\^ea et~al.}{2022}]{deGouvea:2022dtw}
de Gouv\^ea A.,  Martinez-Soler I.,  Perez-Gonzalez Y.~F.,   Sen M.,  2022,
  ArXiv e-print

\bibitem[\protect\citeauthoryear{de Gouvêa, Martinez-Soler, Perez-Gonzalez  \&
  Sen}{de~Gouvêa et~al.}{2020}]{DeGouvea2020}
de Gouvêa A.,  Martinez-Soler I.,  Perez-Gonzalez Y.~F.,   Sen M.,  2020,
  \mn@doi [Physical Review D] {10.1103/physrevd.102.123012}, 102

\bibitem[\protect\citeauthoryear{{van Dokkum} \& {Conroy}}{{van Dokkum} \&
  {Conroy}}{2010}]{2010Natur.468..940V}
{van Dokkum} P.~G.,  {Conroy} C.,  2010, \mn@doi [{Nature}]
  {10.1038/nature09578}, \href
  {https://ui.adsabs.harvard.edu/abs/2010Natur.468..940V} {468, 940}

\bibitem[\protect\citeauthoryear{van Son et~al.,}{van Son
  et~al.}{2021}]{vanSon:2021zpk}
van Son L. A.~C.,  et~al., 2021, ArXiv e-print

\makeatother
\end{thebibliography}

\appendix

\section{Supernova Neutrino Spectra}
\label{app:neutrino}
In order to estimate the DSNB flux, we model the CCSN population by relying on the outputs of one-dimensional, hydrodynamic supernova simulations with Boltzmann neutrino transport from the Garching group~\citep{Garc:SN}. Following \citet{Moller:2018kpn},  three reference CCSN models are used to account for the variations in neutrino emission depending on the mass and fate of the progenitor star. For CCSNe leading to the formation of a neutron star as the compact object remnant, we use models with initial masses of $9.6$ and $27\, \Msun$, whereas for stellar collapses leading to the formation of  black holes, we use the 40\,$\Msun$ ``low'' mass accretion rate model~\citep{Mirizzi:2015eza}. In all  three models, the nuclear equation of state is assumed to be that of Lattimer and Swesty, with a nuclear incompressibility modulus $K = 220$~MeV (LS220 EoS)~\citep{Lattimer:1991nc}. 

The related neutrino energy distributions  are  well described by a pinched Fermi-Dirac distribution~\citep{Keil:2002in, Keil:2003sw, Tamborra:2012ac}:
 \begin{equation}\label{eq:nu_spec}
   \left(\frac{\mathrm{d}n}{\mathrm{d}E}\right)_{\bar\nu_e} = E_\nu^{\rm tot} \frac{\left(1+a\right)^{1+a}}{\Gamma(1+a)} \frac{E^{a}}{\langle E_\nu\rangle^{2+a}} e^{\left[ -\left(1+a\right) \frac{E}{\langle E_\nu \rangle} \right]} \,.
 \end{equation}
Here, the parameters $E_\nu^{\rm tot}$, $a$, and $\langle E_\nu \rangle$ represent the total energy emitted in (anti-electron) neutrinos; a parameter that describes the spectral shape, related to the pinching parameter; and the average energy of the emitted neutrinos, respectively. 

\begin{figure}
     \centering
     \includegraphics[width=\columnwidth]{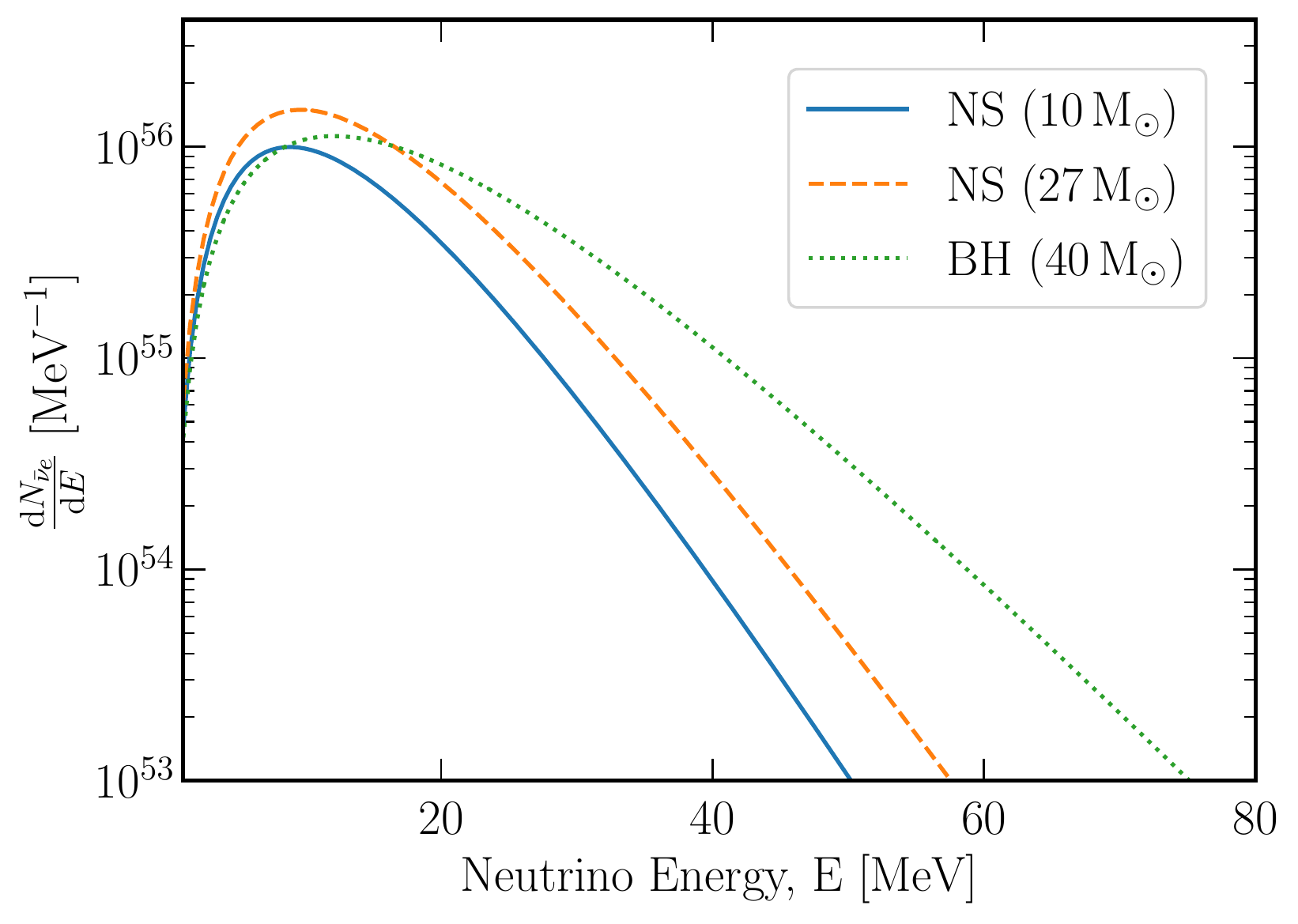}
     \caption{\textbf{Electron antineutrino energy spectra at the source:} Here we show the time-integrated anti-electron neutrino emission spectra of the three reference CCSN models that we use in this work. The blue solid and orange dashed curves correspond to supernovae that result in neutron stars, whose stellar progenitors have $9.7$ and $27\, \Msun$ masses, respectively. The green dotted curve represents a supernova that results in the formation of a black hole and whose stellar progenitor has a mass of $40\, \Msun$. Note that these are the spectra at the source; when observed, these spectra will be redshifted by a factor $1+z$, so that supernovae at high redshift only have non-negligible contributions to the observed spectra at low energies.}
     \label{fig:spectra_at_source}
 \end{figure}

%
While the mean energy, luminosity (energy emitted per time), and pinching  parameter are time-dependent quantities, we are 
interested in  the 
time-integrated neutrino energy distributions. Therefore, we report the time-integrated  characteristic quantities  in Table~\ref{table:fitparams} for all three models adopted to model the DSNB and show these energy spectra in Fig.~\ref{fig:spectra_at_source}. Note that these simulations do not include the effects of neutrino flavor mixing, which we do not take into account  throughout this work.
\begin{table}
 \renewcommand{\arraystretch}{1.3}
   \begin{center}
     \label{tab:specparams}
     \begin{tabular}{c|c|c|c} 
       & $E_\nu^{\rm tot} \,[\mathrm{MeV}]$ & $a$ & $\langle E_\nu\rangle\,[\mathrm{MeV}]$ \\
       \hline
       \hline
       Neutron Star ($10\mathrm{\,M}_{\odot}$) & $1.890\times10^{58}$ & 2.355 & 12.620\\
       Neutron Star ($27\mathrm{\,M}_{\odot}$) & $3.435\times10^{58}$ & 2.307 & 13.856\\
       Black Hole ($40\mathrm{\,M}_{\odot}$) & $4.426\times10^{58}$ & 2.083 & 17.943\\
     \end{tabular}
     \label{table:fitparams}
     \caption{Best fit parameters for the pinched Fermi-Dirac distribution used to describe the numerically generated time-integrated neutrino spectra.}
   \end{center}
\end{table}

\bsp
\label{lastpage}
\end{document}